\documentclass[aps,pra,reprint,onecolumn,amsmath,amssymb,amsfonts,showkeys,showpacs,preprintnumbers]{revtex4-1}
\usepackage{graphicx,xcolor,bm,multirow,lineno,subfigure,rotating,hyperref}

\newcommand{\mc}{\multicolumn}
\newcommand{\mr}{\multirow}
\newcommand{\ia}{\int_{-\infty}^\infty}
\newcommand{\io}{\int_0^\infty}

\DeclareMathOperator{\erf}{erf}

\begin{document}

\title{Distribution of $\mathbf{r}_{12}\cdot\mathbf{p}_{12}$ in quantum systems}
\author{Yves A. Bernard}
\author{Pierre-Fran\c{c}ois Loos}
\email{loos@rsc.anu.edu.au}
\author{Peter M. W. Gill}
\thanks{Corresponding author}
\email{peter.gill@anu.edu.au}
\affiliation{Research School of Chemistry, Australian National University, Canberra, ACT 0200, Australia}

\date{\today}

\begin{abstract}
We introduce the two-particle probability density $X(x)$ of $x=\bm{r}_{12}\cdot\bm{p}_{12}=\left(\bm{r}_1-\bm{r}_2\right) \cdot \left(\bm{p}_1-\bm{p}_2\right)$.  We show how to derive $X(x)$, which we call the Posmom intracule, from the many-particle wavefunction.  We contrast it with the Dot intracule [Y.~A.~Bernard, D.~L.~Crittenden, P.~M.~W.~Gill, Phys.~Chem.~Chem.~Phys., 10, 3447 (2008)] which can be derived from the Wigner distribution and show the relationships between the Posmom intracule and the one-particle Posmom density [Y.~A.~Bernard, D.~L.~Crittenden, P.~M.~W.~Gill, J.~Phys.~Chem.~A, 114, 11984 (2010)].  To illustrate the usefulness of $X(x)$, we construct and discuss it for a number of two-electron systems.
\end{abstract}

\maketitle

\section{\label{sec:intro}Introduction}
Intracules are two-particle density distributions obtained from the spinless second-order reduced density matrix \cite{DavidsonBook}
\begin{equation}
\label{rho2}
	\rho_2
	\begin{pmatrix}
	\bm{r}_1	&,&	\bm{r}_1^{\prime}	\\
	\bm{r}_2	&,&	\bm{r}_2^{\prime}
	\end{pmatrix}
	= \int\Psi^{\ast}\left(\bm{r}_1,\bm{r}_2,\bm{r}_3,\ldots,\bm{r}_N\right)
	\Psi\left(\bm{r}_1^{\prime},\bm{r}_2^{\prime},\bm{r}_3,\ldots,\bm{r}_N\right)d\bm{r}_3\cdots d\bm{r}_N,
\end{equation}
where $\Psi(\bm{r}_1,\bm{r}_2,\bm{r}_3,\ldots,\bm{r}_N)$ is the $N$-particle position wave function.  Intracules are usually normalized to the number of particle pairs $N(N-1)/2$.

The seminal intracule is the Position intracule
\begin{equation}
\label{Pu}
	P(u) = \int \rho_2
	\begin{pmatrix}
	\bm{r}		&,&	\bm{r}		\\
	\bm{r}+\bm{u}	&,&	\bm{r}+\bm{u}
	\end{pmatrix}
	d\bm{r}\,d\bm{\Omega}_{\bm{u}},
\end{equation}
which was introduced long ago by Coulson and Neilson \cite{Coulson61} to study correlation effects in the helium atom.  In \eqref{Pu}, $\bm{u} = \bm{r}_1 - \bm{r}_2$, $u = | \bm{u} | \equiv r_{12}$ and $\bm{\Omega}_{\bm{u}}$ is the angular part of $\bm{u}$. $P(u)$ gives the probability density for finding two particles separated by a distance $u$ and has been widely studied \cite{Curl65, Benesch71, Thakkar76, Thakkar77, Dahl88, Perdew92, Koga93, Cioslowski98, Pu99, Ugalde99, Insights00, Hookium03, Overview03, Omega06, BasisI09, Pearson09, LoosHook, LoosConcentric, Proud13}. 

The corresponding Momentum intracule \cite{Banyard78, Mv02} is
\begin{equation}
\label{Mv}
	M(v) = \frac{1}{(2\pi)^{3}} \int \rho_2
	\begin{pmatrix}
	\bm{r}			&,&	\bm{r}+\bm{q}	\\
	\bm{r}+\bm{u}+\bm{q}	&,&	\bm{r}+\bm{u}
	\end{pmatrix}
	e^{i\bm{q}\cdot\bm{v}}d\bm{r}\,d\bm{q}\,d\bm{u}\,d\bm{\Omega}_{\bm{v}},
\end{equation}
where $\bm{v} = \bm{p}_1 - \bm{p}_2$, $v = | \bm{v} | \equiv p_{12}$ and $\bm{\Omega}_{\bm{v}}$ is the angular part of $\bm{v}$. $M(v)$ gives the probability density for finding two particles moving with a relative momentum $v$.

Starting with the Wigner distribution \cite{Wigner32, Hillery84}, one can construct a family of intracules \cite{Overview03, Omega06}, which provide two-electron position and/or momentum information.  Within this family, the patriarch is the Omega intracule
\begin{equation}
\label{Ouvw}
	\Omega(u,v,\omega)=\frac{1}{(2\pi)^{3}}\int \rho_2
	\begin{pmatrix}
	\bm{r}			&,&	\bm{r}+\bm{q}	\\
	\bm{r}+\bm{u}+\bm{q}	&,&	\bm{r}+\bm{u}
	\end{pmatrix}
	e^{i\bm{q}\cdot\bm{v}}\delta(\omega-\theta_{uv})d\bm{r}\,d\bm{q}\,d\bm{\Omega}_{\bm{u}}\,d\bm{\Omega}_{\bm{v}},
\end{equation}
where $\omega \equiv \theta_{uv}$ is the dynamical angle between the vector $\bm{u}$ and $\bm{v}$.  $\Omega(u,v,\omega)$ can be interpreted as the \emph{joint quasi}-probability density for $u$, $v$ and $\omega$. The \emph{quasi} prefix emphasizes that $\Omega(u,v,\omega)$ is not a rigorous probability density and, indeed, it may take negative values \cite{Wigner32}. Based on the observation of Rassolov \cite{Rassolov99} that both relative position and relative momentum are important to describe the correlation between pairs of electrons, and because the Omega intracule contains information on both quantities, $\Omega(u,v,\omega)$ has been extensively used in Intracule Functional Theory (IFT) \cite{Omega06, OmegaComp07, Sin3w07, AnalKer07, Pars08, BasisI09, RecuRela10, AnnuRep11, PopelierBook}. 

Appropriate integrations \cite{Omega06} reduce the Omega intracule to lower-order intracules such as $P(u)$, $M(v)$ and the Angle intracule \cite{OmegaComp07, Sin3w07} 
\begin{equation}
\label{Uw}
	\Upsilon(\omega)=\io\!\!\!\io\Omega(u,v,\omega)du\,dv,
\end{equation}
which provides information on the angle $\omega$ between $\bm{u}$ and $\bm{v}$.  A similar reduction yields the Dot intracule \cite{Pars08,BasisI09}
\begin{equation}
\label{Dx}
	D(x)=\io\!\!\!\int_{x}^{\infty}\frac{\Omega(u,z/u,\omega)}{u\,z \sin \omega}dz\,du.
\end{equation}
The variable $x = \bm{u}\cdot \bm{v} = u\,v\cos\omega$ combines information on the relative position and momentum of the particles, and it is easy to show that it gives the rate of change of $u^2$, \textit{i.e.}
\begin{equation}
	x = \frac{1}{2} \frac{d}{dt} u^2.
\end{equation}
In this way, $x$ sheds light on the motion of the electrons.  For example, $x = 0$ implies that the electrons are moving in such a way that their separation is constant.  This could arise, for example, if they were in a circular orbit around their centre of mass.

Although $D(x)$ is usually a non-negative function and has proven useful for understanding electronic behaviour \cite{OmegaComp07} and for estimating electron correlation energies in atomic and molecular systems \cite{AnalKer07, Pars08}, its connection to the Omega suggests that it is not a rigorous probability density.  However, in the following Section, we show how to derive the \emph{exact} probability distribution of $x$. 

\section{\label{sec:Th}The Posmom intracule}
We define the Posmom intracule $X(x)$ to be the exact probability density for the variable $x = \bm{u} \cdot \bm{v}$.  It is the two-particle version of the Posmom density $S(s)$ where $s = \bm{r}\cdot\bm{p}$ \cite{Posmom09, PMLett10, PosmomAtom10} and, as we have argued that $s$ describes particle trajectories, we now propose that $x$ likewise characterizes pair trajectories.

The quantum mechanical operator 
\begin{equation}
\label{Os}
	\bar{s}=-i\hbar\left(\frac{3}{2}+\bm{r}\cdot\bm{\nabla}_{\bm{r}}\right)
\end{equation}
is known to be an unbounded self-adjoint operator
\cite{Wess60,Amrein09} and its two-particle equivalent is
\begin{equation}
\label{Ox}
	\bar{x}=-2i\hbar\left(\frac{3}{2}+\bm{u}\cdot\bm{\nabla_u}\right),
\end{equation}
where $\bm{\nabla}$ is the gradient operator. 
Both $\bar{s}$ and $\bar{x}$ correspond to quantum mechanical observables.

Following the same approach used in~\cite{Posmom09} to obtain $S(s)$ from $\bar{s}$, 
one can show that the Posmom intracule can be expressed as the Fourier transform 
\begin{equation}
\label{XxFT}
	X(x)=\frac{1}{2\pi}\ia \widehat{X}(k)e^{ikx}dk
\end{equation}
of the two-particle hyperbolic autocorrelation function
\begin{equation}
\label{Xkrho2}
	\widehat{X}(k)=
	\int \rho_2
	\begin{pmatrix}
	\bm{r}			&,&	\bm{r}+\sinh (k\hbar)\bm{u}\\
	\bm{r}+e^{k\hbar}\bm{u}	&,&	\bm{r}+\cosh (k\hbar)\bm{u}
	\end{pmatrix}
	d\bm{r}d\bm{u}.
\end{equation}
This expression can be simplified after defining the intracule density matrix
\begin{equation}
	\rho_u
	\begin{pmatrix}
	\bm{u}		&,&
	\bm{u}^{\prime}
	\end{pmatrix}
	=\int\rho_2
	\begin{pmatrix}
	\bm{U}-\bm{u}/2		&,&	\bm{U}-\bm{u}^{\prime}/2	\\
	\bm{U}+\bm{u}/2		&,&	\bm{U}+\bm{u}^{\prime}/2
	\end{pmatrix}
	d\bm{U},
\end{equation}
where $\bm{U}=\bm{r}_1+\bm{r}_2$ is the extracule vector, and yields
\begin{equation}
\label{Xkrhou}
	\widehat{X}(k)=\int \rho_u
	\begin{pmatrix}
	e^{+k \hbar} \bm{u}	&,&
	e^{-k \hbar}\bm{u}
	\end{pmatrix}
	d\bm{u}.
\end{equation}

In an entirely analogous way, the Dot intracule can be expressed as the Fourier transform
\begin{equation}
\label{DxFT}
	D(x)=\frac{1}{2\pi}\ia \widehat{D}(k)e^{ikx}dk
\end{equation}
of the $f$-Dot function \cite{Pars08}
\begin{equation}
\label{Dk}
	 \widehat{D}(k)=\int\rho_2
	\begin{pmatrix}
	\bm{r}				&,&	\bm{r}+k\hbar\bm{u}	\\
	\bm{r}+\bm{u}+k\hbar\bm{u}	&,&	\bm{r}+\bm{u}
	\end{pmatrix}
	d\bm{r}\,d\bm{u},
\end{equation}
and the latter can be reduced to
\begin{equation}
\label{Dkrhou}
	 \widehat{D}(k)=\int \rho_u
	\begin{pmatrix}
	(1+k \hbar) \bm{u}	&,&
	(1-k \hbar)\bm{u}
	\end{pmatrix}
	d\bm{u}.
\end{equation}

Comparing \eqref{Xkrhou} with \eqref{Dkrhou} and the Taylor expansion of the exponential function
\begin{equation}
\label{Exp}
	e^{\pm k \hbar}=1\pm k \hbar + \frac{k^2}{2} \hbar^2 + \ldots
\end{equation}
reveals that the probability density $D(x)$ derived from the Wigner distribution is a first-order approximation to the exact density $X(x)$. Thus, the quasi-intracule is correct to $O(\hbar)$ and becomes exact in the classical limit $\hbar\rightarrow 0$.  Remarkably, one can construct the exact density from the approximate density using the mapping
\begin{equation}
\label{XfD}
	\widehat{X}(k) = \frac{\widehat{D}(\tanh(k\hbar))}{\cosh^3(k\hbar)}.
\end{equation}

\begin{table*}
\caption{
\label{tab:DvsI}
One- and two-particle hyperbolic autocorrelation functions.}
\begin{ruledtabular} 
\begin{tabular}{lll}
   			& 	\mc{1}{c}{Density}		&	\mc{1}{c}{Intracule}	
			\\[4pt]
\hline
Dot			&	$\displaystyle\widehat{S}_\text{W}(k)
				=\int\rho_1
				\begin{pmatrix}
				(1+k\hbar/2)\bm{r}	&,&
				(1-k\hbar/2)\bm{r}
				\end{pmatrix}d\bm{r}$
			&	$\displaystyle\widehat{D}(k)=\int\rho_u
				\begin{pmatrix}
				(1+k\hbar)\bm{u}	&,&
				(1-k\hbar)\bm{u}
				\end{pmatrix}d\bm{u}$
				\\[8pt]
Posmom			&	$\displaystyle\widehat{S}(k)
				=\int\rho_1
				\begin{pmatrix}
				e^{+k\hbar/2}\bm{r}	&,&
				e^{-k\hbar/2}\bm{r}
				\end{pmatrix}d\bm{r}$	
			&	$\displaystyle\widehat{X}(k)
				=\int\rho_u
				\begin{pmatrix}
				e^{+k\hbar}\bm{u}	&,&
				e^{-k\hbar}\bm{u}
				\end{pmatrix}d\bm{u}$
				\\[8pt]
\mr{2}{*}{Relation}	
    			&	$\displaystyle\widehat{S}(k)=\frac{\widehat{S}_\text{W}\left(2\tanh(k\hbar/2)\right)}{\cosh^3\left(k\hbar/2\right)}$
			&	$\displaystyle\widehat{X}(k)=\frac{\widehat{D}\left(\tanh(k\hbar)\right)}{\cosh^3\left(k\hbar\right)}$	
			\\[8pt]
			&	$\displaystyle\phantom{\widehat{S}(k)}=\widehat{S}_\text{W}(k) + O(\hbar^2)$	
			&	$\displaystyle\phantom{\widehat{X}(k)}=\widehat{D}(k)+ O(\hbar^2)$								
			\\[8pt]
\end{tabular}
\end{ruledtabular}
\end{table*}

Table \ref{tab:DvsI} gives the one- and two-particle hyperbolic autocorrelation functions, the corresponding first-order (Wigner) approximations, and the relations between them.  In Table \ref{tab:DvsI}, the one-particle density matrix is given by
\begin{equation}
\label{rho1}
	\rho_1
	\begin{pmatrix}
	\bm{r}_1		&,&
	\bm{r}_1^{\prime}
	\end{pmatrix}
	= \frac{2}{N-1} \int 
	\rho_2
	\begin{pmatrix}
	\bm{r}_1	&,&	\bm{r}_1^{\prime}	\\
	\bm{r}_2	&,&	\bm{r}_2
	\end{pmatrix}
	d\bm{r}_2.
\end{equation}
 
If the wave function is expanded in one-electron functions, $\phi_a(\bm{r})$, the reduced two-particle density matrix becomes
\begin{equation}
	\rho_2
	\begin{pmatrix}
	\bm{r}_{1}	&,&	\bm{r}^{\prime}_{1}	\\
	\bm{r}_{2}	&,&	\bm{r}^{\prime}_{2}
	\end{pmatrix}
	=\sum_{abcd} P_{abcd}\phi_a(\bm{r}_1)\phi_b(\bm{r}^{\prime}_1)\phi_c(\bm{r}_2)\phi_d(\bm{r}^{\prime}_2),
\end{equation}
where $P_{abcd}$ is a two-particle density matrix element. In this case, \eqref{Xkrho2} is given by 
\begin{equation}
\label{xkabcd}
	\widehat{ X }(k)=\sum_{abcd} P_{abcd}\left[abcd\right]_{\widehat{X}},
\end{equation}
where we have introduced the two-particle hyperbolic autocorrelation integral $\left[abcd\right]_{\widehat{X}}$.  For example, if the basis functions are $s$-type Gaussians, we obtain (in atomic units)
\begin{align}
\label{xkssss}
	\left[ssss\right]_{\widehat{X}}
	& = \int e^{-\alpha|\bm r-\bm A|^2} e^{-\beta|\bm r+\sinh (k)\bm u-\bm B|^2} e^{-\gamma|\bm r+\exp (k)\bm u-\bm C|^2} e^{-\delta|\bm r+\cosh (k) \bm u-\bm D|^2}\,d\bm r\,d\bm u	\notag	\\
	& = \frac{\pi^3}{J^{3/2}} \exp\left[\frac{1}{\xi}\Big(\frac{|\bm{H}|^2}{J }-F\Big)\right],
\end{align}
where
\begin{subequations}
\begin{gather}
	\xi=\alpha+\beta+\gamma+\delta,\\
	J=\xi[\beta\sinh^2(k)+\gamma \exp^2(k)+\delta\cosh^2(k)]-[\beta\sinh(k)+\gamma \exp(k)+\delta\cosh(k)]^2,
\\
	F=\alpha \beta |\bm{A}-\bm{B}|^2+\alpha \gamma |\bm{A}-\bm{C}|^2+\alpha \delta |\bm{A}-\bm{D}|^2 + \beta \gamma |\bm{B}-\bm{C}|^2+ \beta \delta |\bm{B}-\bm{D}|^2+ \gamma \delta |\bm{C}-\bm{D}|^2,
\end{gather}
\end{subequations}
and
\begin{subequations} \label{eq:Hetc}
\begin{gather}
	\bm{H}		=	\sinh(k) \bm{G_B}+ \exp(k) \bm{G_C}+ \cosh(k) \bm{G_D},
	\\
	\bm{G_B}	=	\alpha \beta (\bm{B}-\bm{A})+\beta \gamma(\bm{B}-\bm{C})+\beta \delta(\bm{B}-\bm{D}),
	\\
	\bm{G_C}	=	\alpha \gamma (\bm{C}-\bm{A})+\beta \gamma(\bm{C}-\bm{B})+\gamma \delta(\bm{C}-\bm{D}),
	\\
	\bm{G_D}	=	\alpha \delta (\bm{D}-\bm{A})+\beta \delta(\bm{D}-\bm{B})+\gamma \delta(\bm{D}-\bm{C}).
\end{gather}
\end{subequations}
The scalars and vectors above are independent of the choice of origin and $\widehat{X}(k)$ and $X(x)$ are therefore likewise independent.  This contrasts with the annoying origin-dependence \cite{PMLett10} of the one-electron posmom density $S(s)$.

Integrals of higher angular momentum can be generated by differentiating $\left[ssss\right]_{\widehat{X}}$ with respect to the Cartesian coordinates of the basis function centers, as first suggested by Boys \cite{Boys50}, or, more efficiently, using recurrence relations \cite{RecuRela10}.  We have written a program to compute $X(x)$ within an $spd$ Gaussian basis set and implemented this in a development version of the \textsc{Q-Chem 3.2} quantum chemistry package \cite{QChem3}.

Eqs \eqref{xkabcd} -- \eqref{eq:Hetc} are easily modified to generate $P(u)$, $M(v)$, $\Upsilon(\omega)$ and $\widehat{D}(k)$.  In particular, if the functions $\sinh(k)$, $\exp(k)$ and $\cosh(k)$ are replaced by their first-order approximations ($k$, $1+k$ and $1$, respectively) in the expressions for $J$ and $\bm{H}$, one obtains the $f$-Dot integrals $[ssss]_{\widehat{D}}$ \cite{Pars08}.

In the special case of concentric $s$-type Gaussians, the intracule integrals become
\begin{subequations}
\begin{align}
	\left[ssss\right]_{P}		&	=	\frac{4\pi^{5/2}}{\xi^{3/2}}u^2 \exp\left(-\frac{\mu}{\xi} u^2\right),
	\label{PuCssss}\\
	\left[ssss\right]_{M}		&	=	\frac{4\pi^{5/2}}{\chi^{3/2}}v^2 \exp\left(-\frac{\nu}{\chi} v^2\right),
	\label{MvCssss}\\
	\left[ssss\right]_{\Upsilon}	&	=	\frac{\pi^3\left(\lambda-2\eta \cos^2\omega\right)}{2 \zeta^{3/2} 
							\left(\lambda+\eta  \cos ^2\omega \right)^{5/2}}\sin \omega,
	\label{UwCssss}\\
	\left[ssss\right]_{\widehat{D}}	&	=	\frac{\pi^3}{K^{3/2}},
	\label{dkCssss}\\
	\left[ssss\right]_{\widehat{X}}	&	=	\frac{\pi^3}{J^{3/2}},
	\label{xkCssss}
\end{align}
\end{subequations}
where
\begin{subequations}
\begin{align}
	\mu		&	=	(\alpha+\beta)(\gamma+\delta),
	\\
	\nu		&	=	(\alpha+\gamma)(\beta+\delta),
	\\
   	\zeta	&	=	(\alpha +\delta ) (\beta +\gamma),
	\\
	\chi		&	=	4(\alpha \beta \gamma+\alpha \beta \delta+\alpha \gamma \delta+\beta \gamma \delta),
	\\
	\lambda	&	=	\left(\frac{1}{\alpha +\delta }+\frac{1}{\beta +\gamma }\right) \left(\frac{\alpha \delta}{\alpha+\delta}+\frac{\beta \gamma}{\beta+\gamma}\right),
	\\
   	\eta		&	=	\left(\frac{\alpha }{\alpha +\delta }-\frac{\beta }{\beta +\gamma }\right)^2,
	\\
	K		&	=	\xi\left[\beta k^2+\gamma (1+k)^2+\delta\right]-\left[\beta k+\gamma (1+k)+\delta\right]^2.
\end{align}
\end{subequations}

In the calculations described below, we have computed $X(x)$ and $D(x)$ numerically using Eqs. \eqref{XxFT} and \eqref{DxFT}.  $\widehat{D}(k)$, $\widehat{X}(k)$, $D(x)$ and $X(x)$ are all even functions and we will therefore focus only on $x\geq 0$ and $k\geq 0$.

\begin{figure*}
	\includegraphics[width=0.80\textwidth]{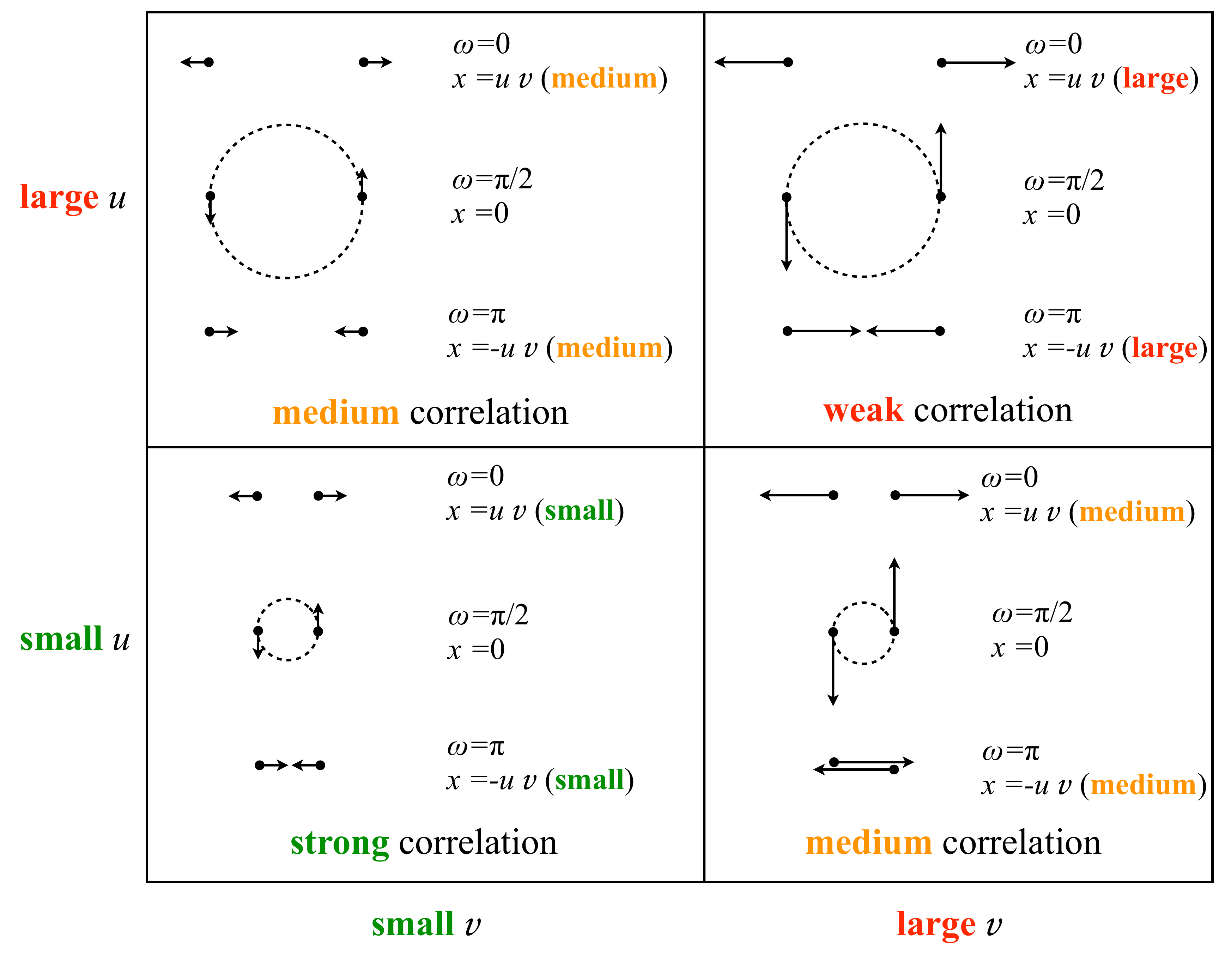}
\caption{
\label{fig:uvw}
Physical interpretation of the variables $u$, $v$, $\omega$ and $x$ in the weak, medium and strong correlation regimes.}
\end{figure*}

Physical interpretations of the variables $u$, $v$, $\omega$ and $x$ are summarized in Fig. \ref{fig:uvw}. The three limiting configurations $\omega = 0$, $\pi/2$ and $\pi$ (which correspond to $x = u\,v$, $0$ and $-u\,v$) are depicted for the weak ($u$ and $v$ large), medium (where one of $u$ and $v$ is large and the other is small) and strong correlation ($u$ and $v$ small) regimes. A faithful description of electron correlation requires information about the relative position $\bm{u}$ and momentum $\bm{v}$, but also on the mutual orientation $\omega$ of these two vectors, which gives insight into the nature of the electrons' mutual orbit.  The Dot and Posmom intracules provide information about the distribution of values of $x = u v \cos\omega$, and thus about the type of correlation regime (weak, medium or strong). However, as noted above, being a first-order approximation of $X(x)$, the information gathered in $D(x)$ is slightly biased. The effects of this approximation will be investigated below.

In Section \ref{sec:Hook}, the Posmom intracule is investigated alongside $D(x)$, $\Upsilon(\omega)$, $P(u)$ and $M(v)$ for the two electrons in a parabolic quantum dot.  In Section \ref{sec:He}, we turn our attention to the electrons in a helium atom or helium-like ion.  We also compare the Posmom intracules for ground and excited states and study the effect of the dimensionality of the space $\mathcal{D}$.  Atomic units are used throughout.

\section{\label{sec:Hook} Parabolic quantum dots}
In our study of the Posmom intracule in parabolic quantum dots \cite{Kestner62}, we consider three different treatments of the Coulomb interaction between the two electrons.  First, the non-interacting case, in which it is simply ignored;  second, the Hartree-Fock (HF) case \cite{Szabo} in which it is approximated in a mean-field sense;  third, the exact treatment which is possible for certain values of the harmonic confinement force constant \cite{Hook2w89, HookAllw93}. 

\subsection{Hamiltonian and wave functions}
The Hamiltonian is
\begin{equation} \label{H0-QD}
	\mathcal{H} = - \frac{1}{2}(\nabla_1^2 + \nabla_2^2) + V(r_1) + V(r_2) + \frac{1}{r_{12}},
\end{equation}
where 
\begin{equation}
	V(r) = \frac{r^2}{2\kappa^2}
\end{equation}
is the external harmonic potential and $1/\kappa^2$ is the force constant. 

The $^1S$ ground state of the non-interacting system has the wave function
\begin{gather}
	\Psi_0(\bm{r}_1,\bm{r}_2)=\psi_0(\bm{r}_1)\psi_0(\bm{r}_2),\\
	\psi_0(\bm{r})=\left(\pi\kappa\right)^{-3/4}\exp \left(-\frac{r^2}{2\kappa}\right),
\end{gather}
and the energy 
\begin{equation}
	E_{0} = \frac{3}{\kappa}.
\end{equation}

The more accurate HF wave function
\begin{equation}
	\Psi_\text{HF}(\bm{r}_1,\bm{r}_2)=\psi_\text{HF}(\bm{r}_1)\psi_\text{HF}(\bm{r}_2)
\end{equation}
is not known in closed form, but can be efficiently treated numerically by expanding $\psi_\text{HF}(\bm{r})$ in a Gaussian basis
\begin{equation}
\label{HF}
	\psi_\text{HF}(\bm{r})=\sum_{j=1}^{N_{\text{G}}}c_j \exp(-\alpha_j r^2).
\end{equation}
The HF energy can be directly minimized with respect to the coefficients $c_j$ and exponents $\alpha_j$ using a numerical solver \cite{Math8}, thus avoiding the self-consistent field procedure usually needed for this kind of calculation \cite{RagotHook08, LoosBall10}.

The exact wave function and energy can be found in closed form \cite{HookAllw93} for certain values of $\kappa$. For example, for $\kappa=2$
\begin{subequations}
\begin{gather}
	\Psi_{2}(\bm{r}_1,\bm{r}_2) = \left(1+\frac{r_{12}}{2} \right) \psi_0(\bm{r}_1) \psi_0(\bm{r}_2),	\\
	E_{2} = 2,
\end{gather}
\end{subequations}
and, for $\kappa=10$,
\begin{subequations}
\begin{gather}
   	\Psi_{10}(\bm{r}_1,\bm{r}_2) = \left(1+\frac{r_{12}}{2}+\frac{r_{12}^2}{20}\right) \psi_0(\bm{r}_1) \psi_0(\bm{r}_2),	\\
	E_{10} = 1/2.
\end{gather}
\end{subequations}

Table \ref{tab:EHk} shows the convergence of $E_\text{HF}$ with $N_\text{G}$ for $\kappa=2$ and $\kappa=10$. 
The correlation energy
\begin{equation}
	E_\text{c} = E_\text{exact} - E_\text{HF}
\end{equation}
for $\kappa=2$ ($E_\text{c} = 38.438\,871\,755$ m$E_{\text{h}}$) agrees with earlier work \cite{Hookium03,RagotHook08}.  For $\kappa=10$, we find $E_\text{c} = 29.041\,525\,56$ m$E_{\text{h}}$.

\begin{table}
\caption{
\label{tab:EHk}
Energies of parabolic quantum dots for different treatments of the interelectronic interaction.}
\begin{ruledtabular} 
\begin{tabular}{ccllll}
						&	$N_{\text{G}}$						& \mc{1}{c}{$\kappa=2$} 		& \mc{1}{c}{$\kappa=10$}		\\
\hline
$E_0$					&	---									& 1.5 							& 0.3							\\
\hline
\mr{7}{*}{$E_{\text{HF}}$}	&	1								& 2.04							& 0.53							\\
						&	2									& 2.038\,439 					& 0.529\,04						\\
						&	3									& 2.038\,438\,9 				& 0.529\,041\,5					\\
						&	4									& 2.038\,438\,871\,8		 	& 0.529\,041\,525\,6			\\
						&	5									& 2.038\,438\,871\,755 			& 0.529\,041\,525\,56			\\
						&	O'Neill and Gill\footnotemark[1]		& 2.038\,438\,87 				& \mc{1}{c}{---}					\\
						&	Ragot\footnotemark[2]				& 2.038\,438\,871\,76 			& \mc{1}{c}{---}					\\
\hline
$E_{\text{exact}}$		&	---									& 2.								& 0.5
\end{tabular}
\end{ruledtabular} 
\footnotetext[1]{Reference \cite{Hookium03}: 7 basis functions.}
\footnotetext[2]{Reference \cite{RagotHook08}: 11 basis functions.}
\end{table}

\subsection{Position Intracule}
The non-interacting Position intracule is
\begin{equation}
	P_0(u)=\sqrt{\frac{2}{\pi\kappa^{3}}}u^2\exp\left(-\frac{u^2}{2\kappa}\right).
\end{equation}
and the HF intracule $P_{\text{HF},\kappa}(u)$ is found from \eqref{PuCssss} and \eqref{HF}.  For $\kappa=2$ and $\kappa=10$, the exact intracules are given by
\begin{gather}
	P_{2}(u)=\frac{(1+u/2)^2}{8+5\sqrt{\pi}} u^2 \exp\left(-\frac{u^2}{4}\right),
	\label{Pu2}\\
	P_{10}(u)=\frac{(1+u/2+u^2/20)^2}{5/2(240+61\sqrt{5\pi})} u^2 \exp\left(-\frac{u^2}{20}\right).
	\label{Pu10}
\end{gather}
Equation \eqref{Pu2} has been reported previously \cite{Hookium03}.

\subsection{Momentum Intracule}
Using the same notation as above, the non-interacting Momentum intracule is
\begin{equation}
	M_0(v) = \frac{2}{\pi}\kappa^{3/2}v^2 \exp\left(-\frac{\kappa v^2}{2}\right),
\end{equation}
$M_{\text{HF},2}(v)$ and $M_{\text{HF},10}(v)$ are obtained from \eqref{MvCssss} and \eqref{HF}, and the exact Momentum intracules are
\begin{gather}
\label{Mv2}
	M_{2}(v) = \frac{8v^2}{8+5\sqrt{\pi}}
	\left[\sqrt{\frac{2}{\pi}}+e^{-\frac{v^2}{2}}+\left(\frac{1}{iv}+iv\right) \erf\left(\frac{iv}{\sqrt{2}}\right) e^{-\frac{v^2}{2}}\right]^2,
\\
\label{Mv10}
	M_{10}(v) = \frac{80\sqrt{5}v^2}{48\sqrt{5}+61\sqrt{\pi}}
	\left[\sqrt{\frac{10}{\pi}}+\left(4-5v^2\right)e^{-\frac{5v^2}{2}}+\left(\frac{1}{iv}+5iv\right) \erf\left(\sqrt{\frac{5}{2}}iv\right)e^{\frac{-5v^2}{2}}\right]^2,
\end{gather}
where $\erf(z)$ is the error function \cite{NISTbook}.  Equation~\eqref{Mv2} has been reported previously \cite{Hookium03}.

\subsection{Angle Intracule}
The Angle intracule of two non-interacting particles is entirely determined by the Jacobian factor and is \cite{Sin3w07}
\begin{equation}\label{Uw0}
	\Upsilon_0(\omega)=\frac{1}{2}\sin\omega.
\end{equation}
$\Upsilon_{\text{HF},2}(\omega)$ and $\Upsilon_{\text{HF},10}(\omega)$ are obtained from \eqref{UwCssss} and \eqref{HF}. $\Upsilon_{2}(\omega)$ and $\Upsilon_{10}(\omega)$ have been obtained by numerical integration of \eqref{Ouvw} and \eqref{Uw}. Eq.~\eqref{Ouvw} can be reduced to a two-dimensional integral, and the resulting four-dimensional numerical integration in Eq.~\eqref{Uw} was performed carefully to ensure accuracy of $O(10^{-3})$ for each value of $\omega$. 

\subsection{Dot and Posmom intracules}
$\widehat{D}_{\text{HF},2}(k)$ and $\widehat{D}_{\text{HF},10}(k)$ [Eqs. \eqref{HF} and \eqref{dkCssss}], as well as $\widehat{X}_{\text{HF},2}(k)$ and $\widehat{X}_{\text{HF},10}(k)$ [Eqs. \eqref{HF} and \eqref{xkCssss}] have been obtained numerically. Table \ref{tab:DxXx} gathers the non-interacting and exact ($\kappa=$ 2 and 10) Dot and Posmom intracules in Fourier and real space. The similarity between the Dot and Posmom expressions is striking.

\subsection{Holes}
\begin{figure}
	\subfigure[Position]{
	\includegraphics[width=0.4\textwidth]{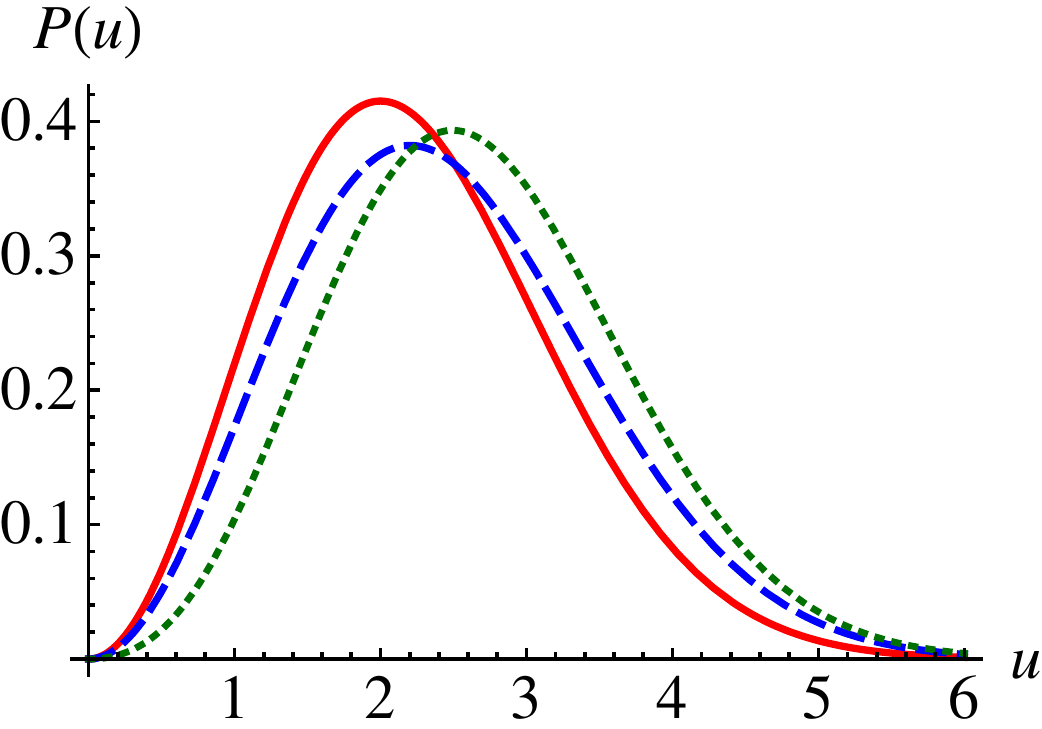}\label{fig:HO_2_Pu}}
	\subfigure[Momentum]{
	\includegraphics[width=0.4\textwidth]{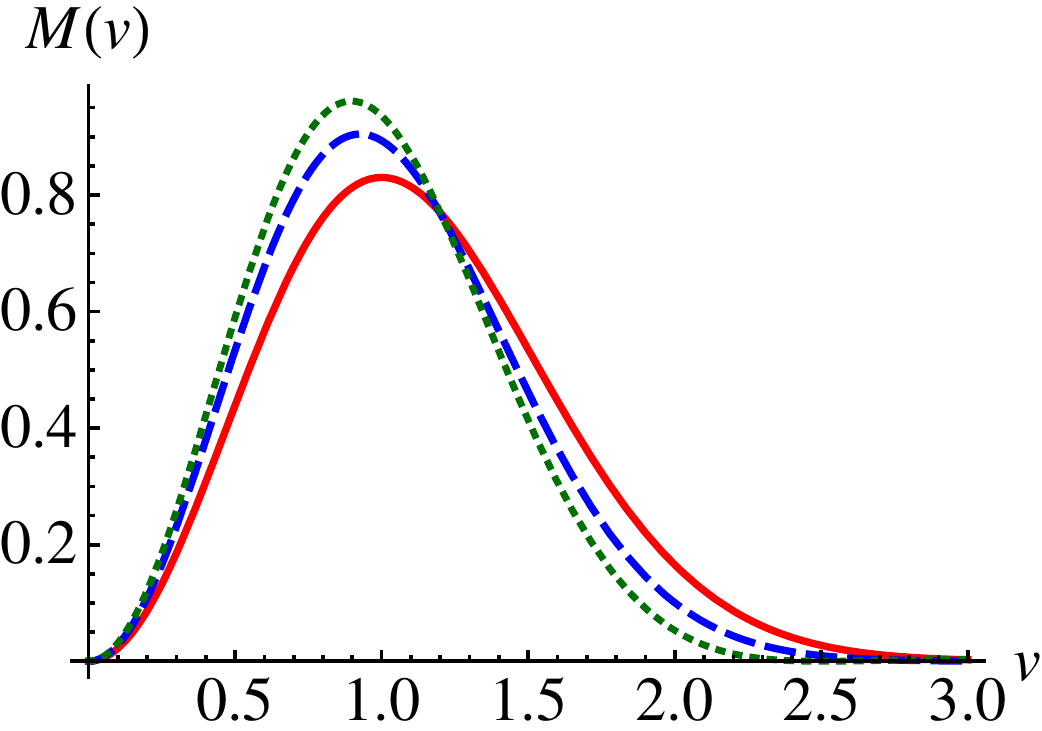}\label{fig:HO_2_Mv}}\\
	\subfigure[Angle]{
	\includegraphics[width=0.4\textwidth]{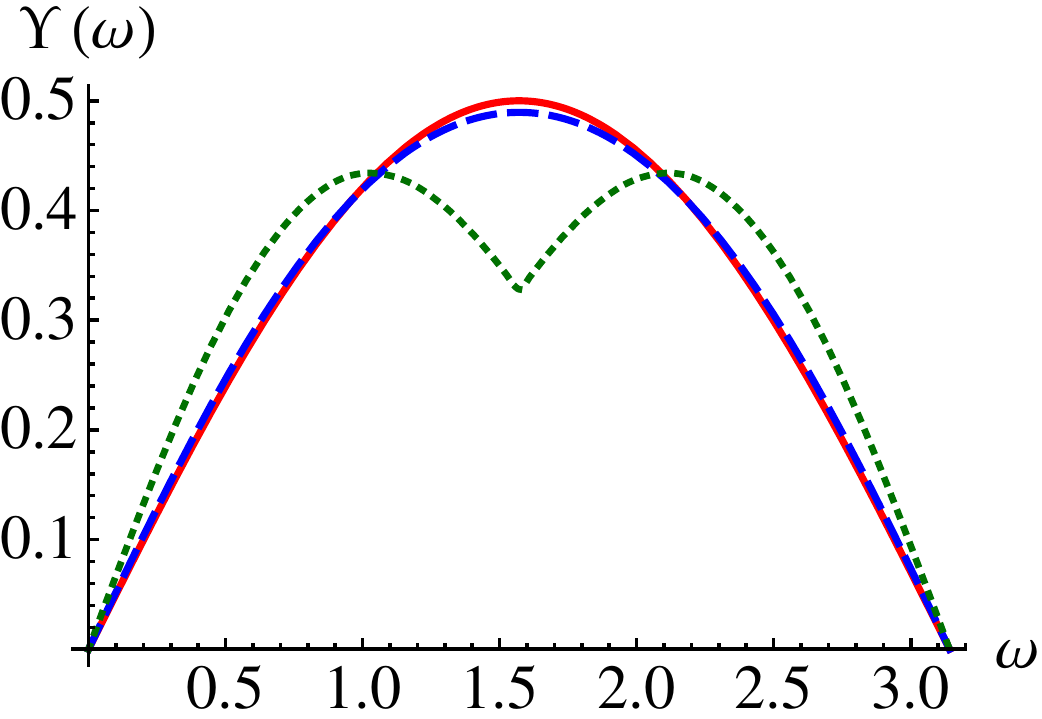}\label{fig:HO_2_Uw}}\\
	\subfigure[Dot]{
	\includegraphics[width=0.4\textwidth]{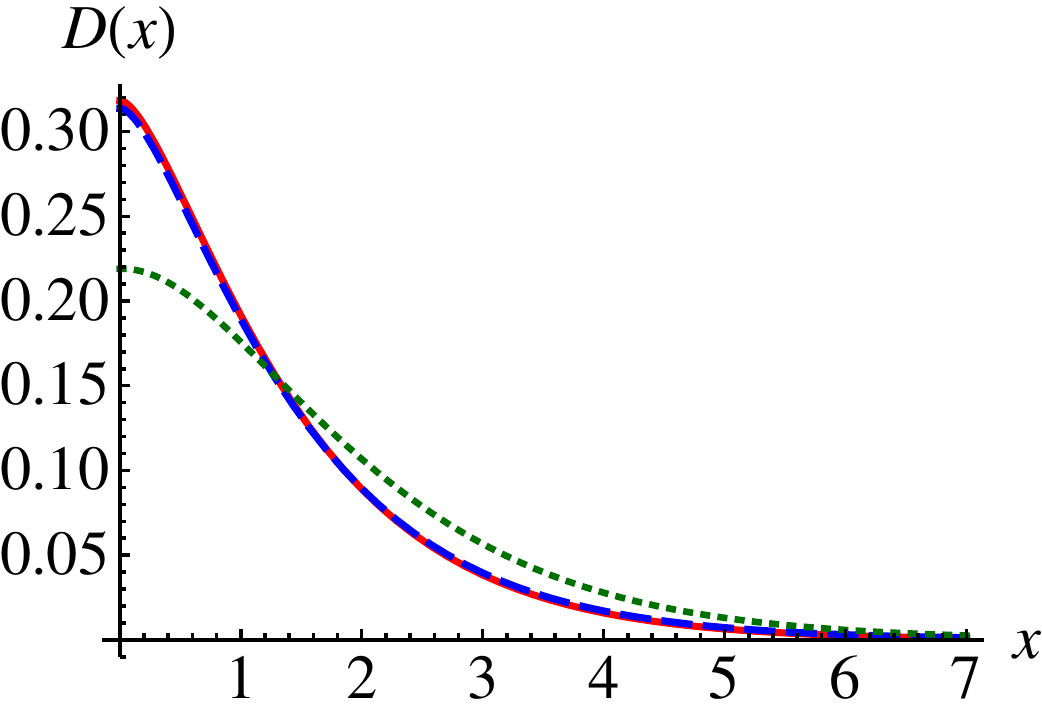}\label{fig:HO_2_Dx}}
	\subfigure[Posmom]{
	\includegraphics[width=0.4\textwidth]{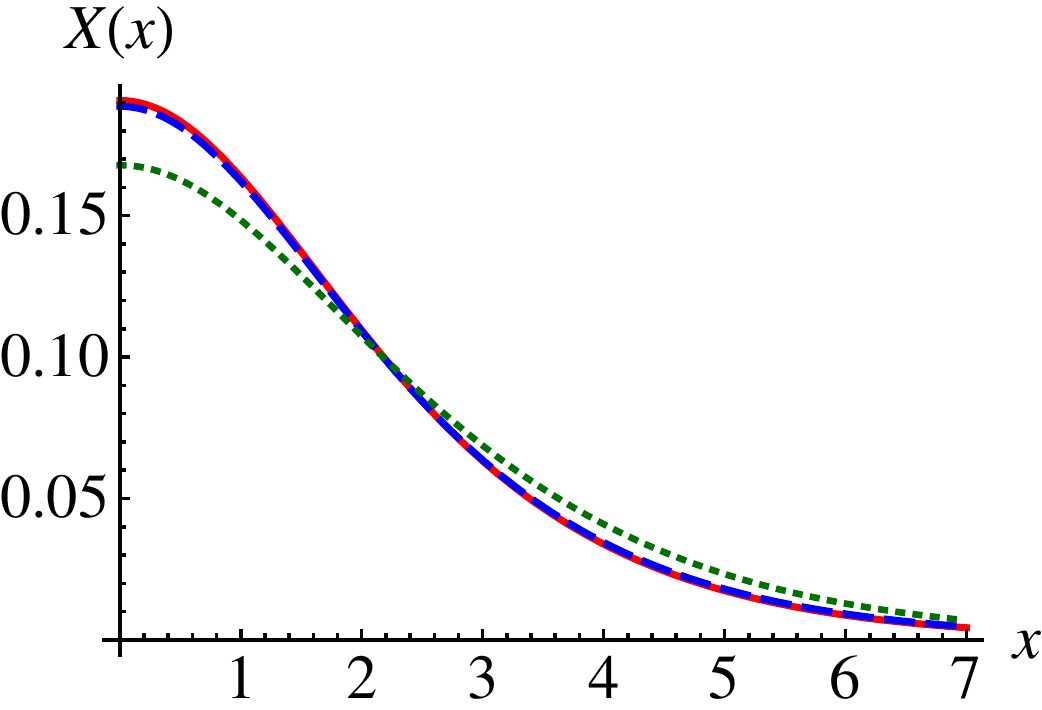}\label{fig:HO_2_Xx}}
\caption{
\label{fig:HO}
Intracules for a parabolic quantum dot with $\kappa=2$: non-interacting(---), HF (- - -) and exact ($\cdots$).}
\end{figure}

\begin{figure}
	\subfigure[Position $\kappa=2$]{
	\includegraphics[width=0.3\textwidth]{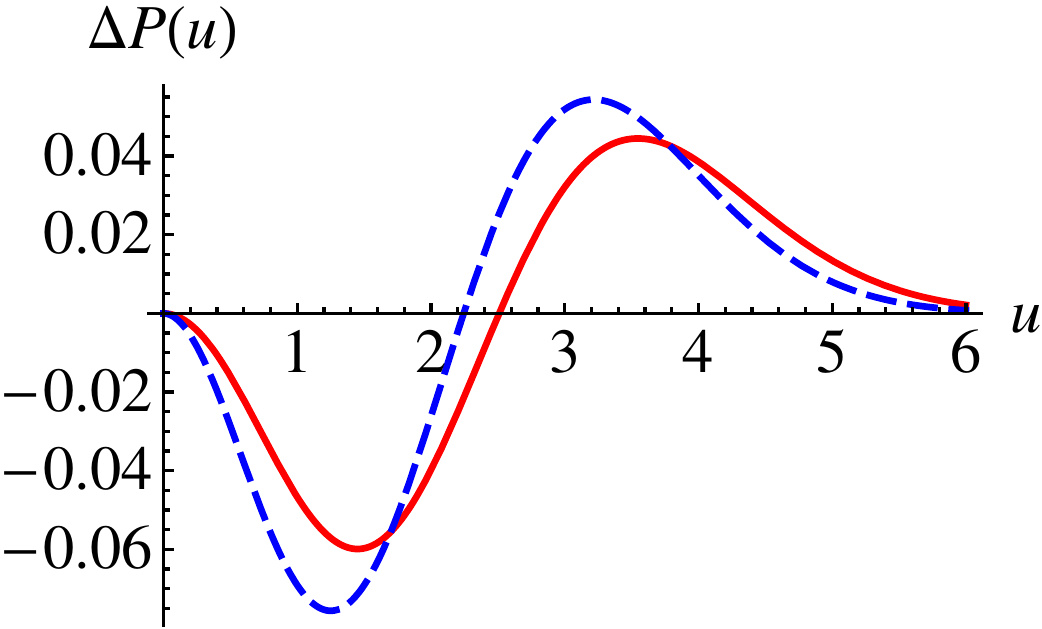}\label{fig:HO_2_DPu}}
	\subfigure[Position $\kappa=10$]{
	\includegraphics[width=0.3\textwidth]{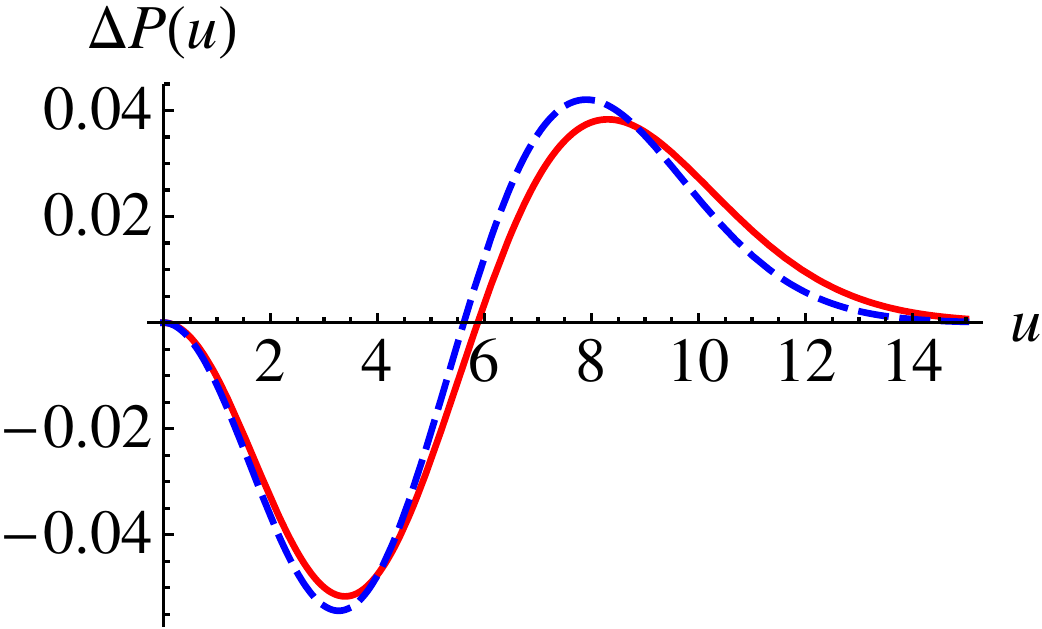}\label{fig:HO_10_DPu}}\\
	\subfigure[Momentum $\kappa=2$]{
	\includegraphics[width=0.3\textwidth]{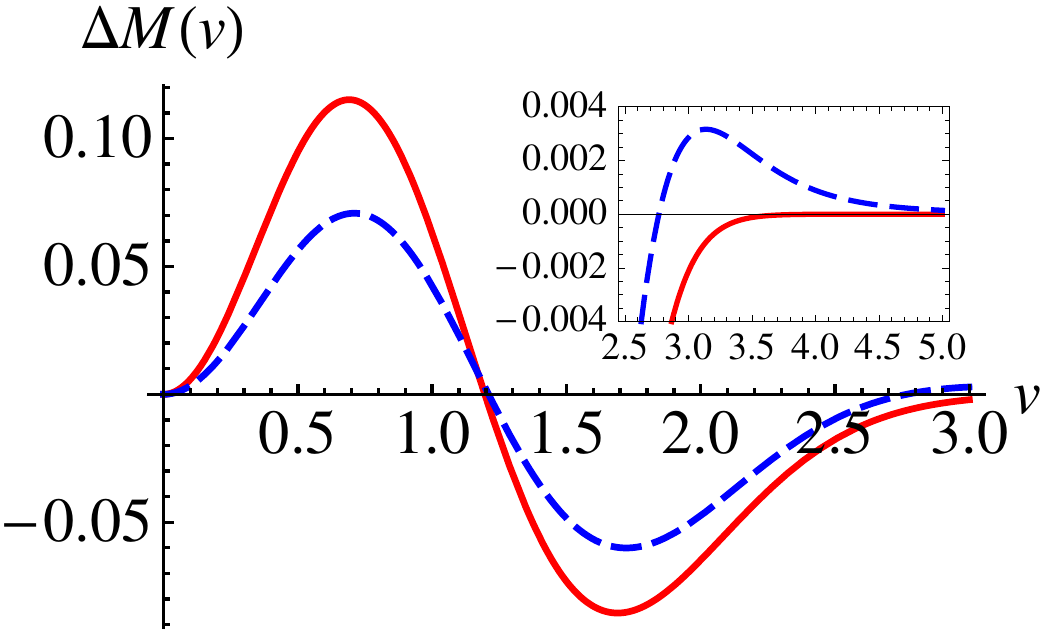}\label{fig:HO_2_DMv}}
	\subfigure[Momentum $\kappa=10$]{
	\includegraphics[width=0.3\textwidth]{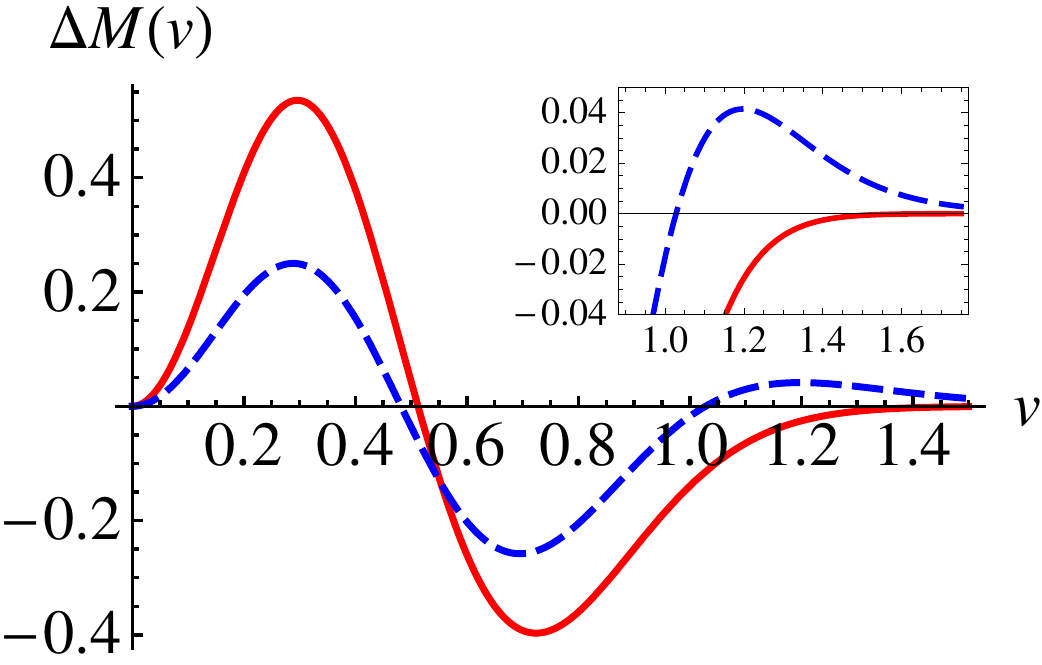}\label{fig:HO_10_DMv}}\\
	\subfigure[Angle $\kappa=2$]{
	\includegraphics[width=0.3\textwidth]{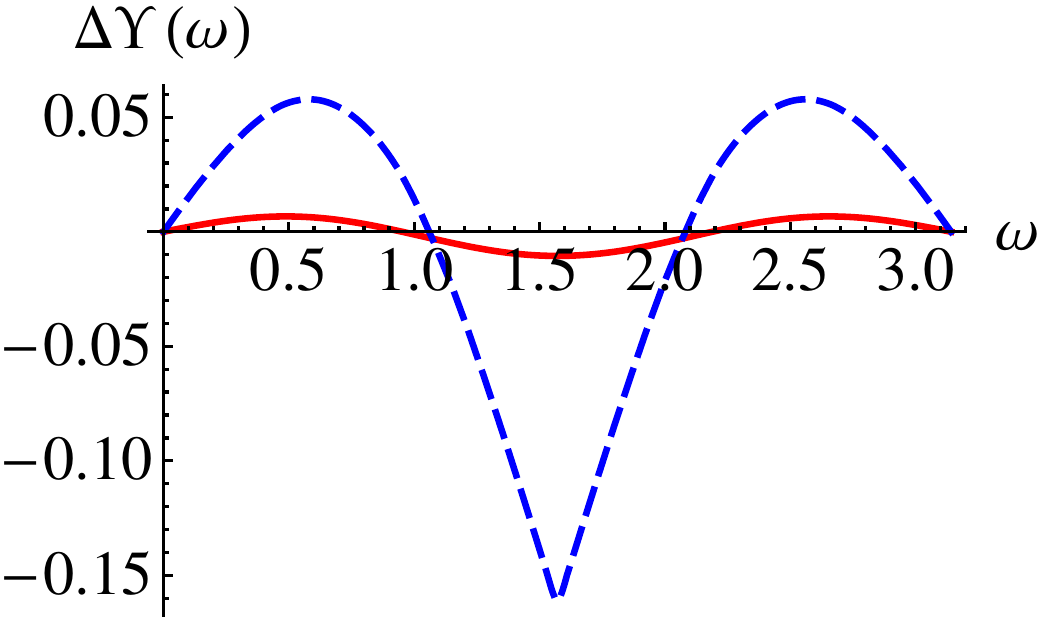}\label{fig:HO_2_DUw}}
	\subfigure[Angle $\kappa=10$]{
	\includegraphics[width=0.3\textwidth]{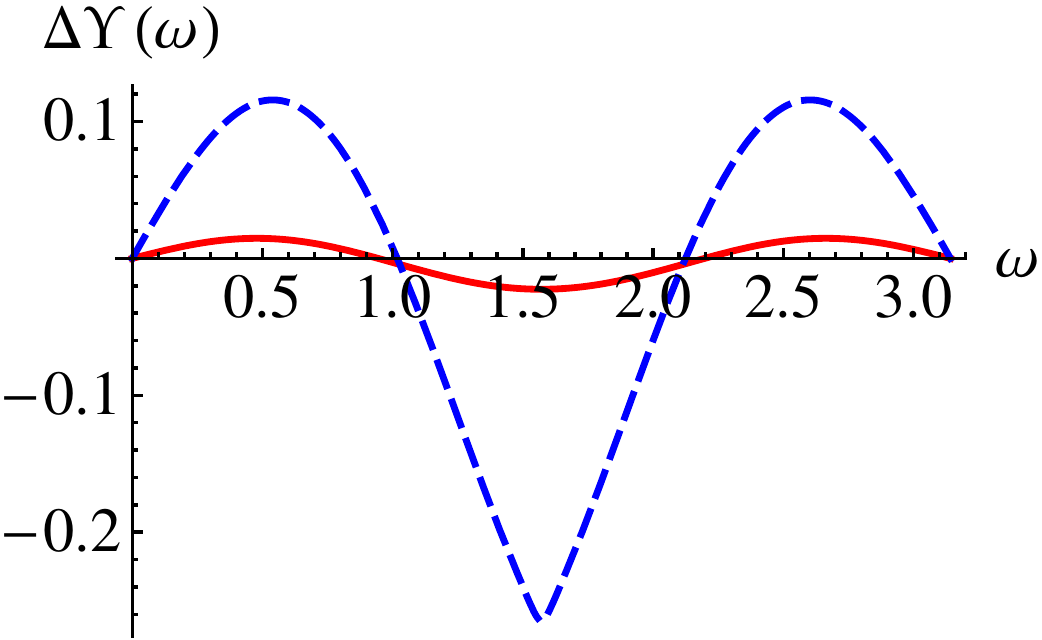}\label{fig:HO_10_DUw}}\\
	\subfigure[Dot $\kappa=2$]{
	\includegraphics[width=0.3\textwidth]{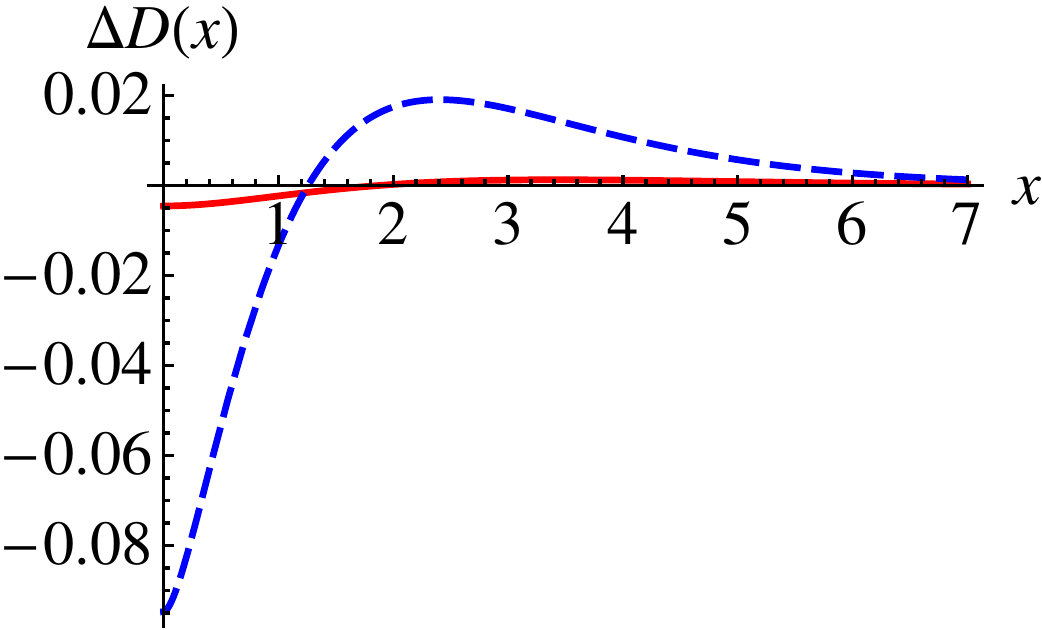}\label{fig:HO_2_DDx}}
	\subfigure[Dot $\kappa=10$]{
	\includegraphics[width=0.3\textwidth]{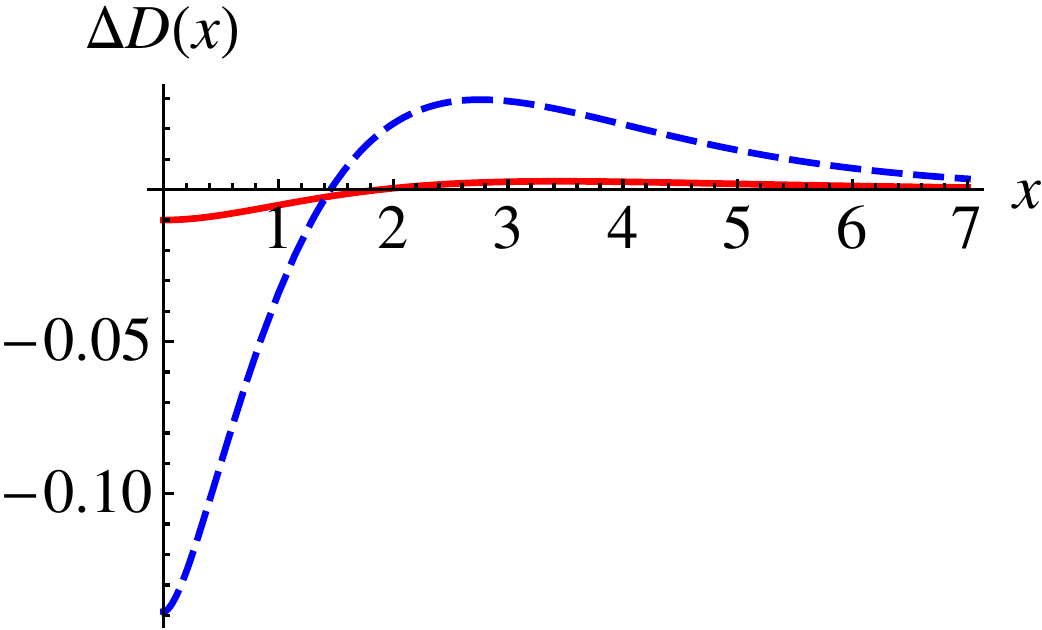}\label{fig:HO_10_DDx}}\\
	\subfigure[Posmom $\kappa=2$]{
	\includegraphics[width=0.3\textwidth]{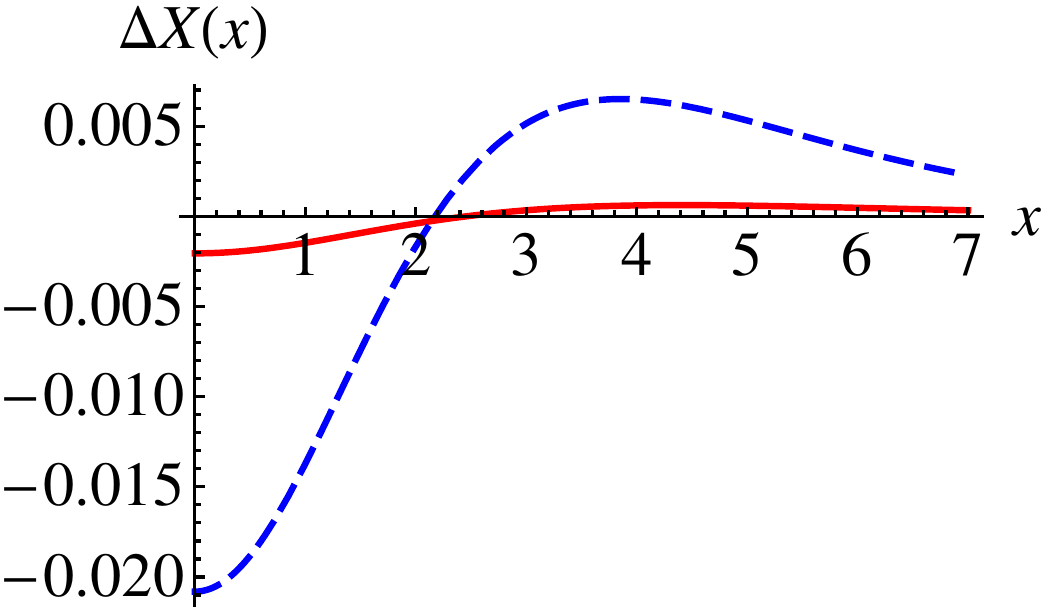}\label{fig:HO_2_DXx}}
	\subfigure[Posmom $\kappa=10$]{
	\includegraphics[width=0.3\textwidth]{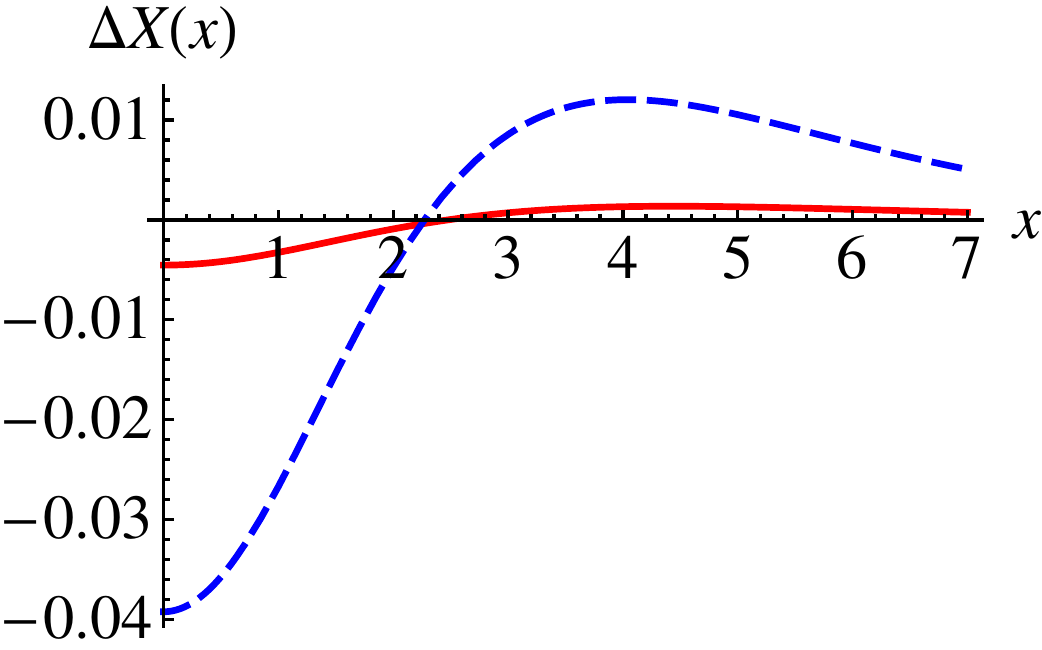}\label{fig:HO_10_DXx}}
\caption{
\label{fig:DHO}
Intracule holes for the parabolic quantum dots with $\kappa=2$ and $\kappa=10$: HF hole (---) and correlation hole (- - -).}
\end{figure}

The correlation hole was originally defined \cite{Coulson61} as the difference between the exact and HF Position intracule
\begin{equation}
	\Delta P(u)= P(u)-P_{\text{HF}}(u),
\end{equation}
but this can be extended to any intracule $I$
\begin{equation} \label{DI}
	\Delta I = I-I_{\text{HF}}.
\end{equation}
One can also define the HF hole as the difference between the HF and non-interacting intracules
\begin{equation}
	\Delta I_\text{HF}= I_{\text{HF}}-I_0.
\end{equation}
Fig. \ref{fig:HO} shows all of the intracules for $\kappa=2$ and Fig.~\ref{fig:DHO} shows the holes created as the Coulomb interaction is introduced.

One can see from $P(u)$ and $M(v)$ in Figs. \ref{fig:HO_2_Pu} or \ref{fig:HO_2_Mv} that the electrons are found at larger separations and move with lower relative momenta in the HF approximation than in the non-interacting case.  However, the non-interacting and HF intracules, $\Upsilon(\omega)$, $D(x)$ and $X(x)$, are almost identical.  The fact that $\Upsilon(\omega)$, $D(x)$ and $X(x)$ are all invariant under a uniform scaling leads us to conclude that the introduction of the Coulomb operator at the mean-field level leads to an almost exact dilation of the system.

\begin{turnpage}
\begin{table*}
\caption{
\label{tab:DxXx}
Dot and Posmom intracules for the non-interacting and the exact treatments in Fourier and real spaces for two parabolic quantum dots.  Note that $\Gamma$ is the gamma function \cite{NISTbook} and $\tilde{K}_n=x^nK_n$, where $K_n$ represents the $n$th modified Bessel function of the second kind \cite{NISTbook}.}
\begin{ruledtabular} 
\small
\begin{tabular}{|c|c|c|c|}
		&	&	Dot	&	Posmom\\[4pt]
\hline
\mr{3}{*}{\rotatebox{90}{Fourier space}}
&
Non-int.
&
$\frac{1}{(1+k^2)^{3/2}}$
&
$\frac{1}{\cosh^{3/2}(2k)}$
\\[8pt]
&
$\kappa=2$
&
$\frac{1}{8+5\sqrt{\pi}}
\left[-\frac{\sqrt{\pi}}{\left(1+k^2\right)^{3/2}}
+\frac{8}{\left(1+k^2\right)^2}
+\frac{6\sqrt{\pi}}{\left(1+k^2\right)^{5/2}}\right]$
&
$\frac{1}{8+5\sqrt{\pi}} 
\left[\frac{2\sqrt{\pi}}{\cosh^{3/2}(2k)}
+\frac{8\cosh(k)}{\cosh^2(2k)}
+\frac{3\sqrt{\pi}}{\cosh^{5/2}(2k)}\right]$
\\[8pt]
&
$\kappa=10$
&
$\frac{1}{240+61 \sqrt{5\pi}}
\left[\frac{\sqrt{5\pi}}{\left(1+k^2\right)^{3/2}}
-\frac{80}{\left(1+k^2\right)^2}
+\frac{320}{\left(1+k^2\right)^3}
+\frac{60\sqrt{5\pi}}{\left(1+k^2\right)^{7/2}}\right]$
&
$\frac{1}{240+61\sqrt{5\pi}} 
\left[\frac{16\sqrt{5\pi}}{\cosh^{3/2}(2k)}
+\frac{80\cosh(k)}{\cosh^2(2k)}
+\frac{30\sqrt{5\pi}}{\cosh^{5/2}(2k)}
+\frac{160\cosh(k)}{\cosh^3(2k)}
+\frac{15\sqrt{5\pi}}{\cosh^{7/2}(2k)}\right]$
\\[8pt]
\hline
\mr{3}{*}{\rotatebox{90}{Real space}}
&
Non-int.
&
$\frac{1}{\pi} x K_1(x)$
&
$\frac{1}{\sqrt{2\pi^3}}
\left|\Gamma\left(\frac{3+ix}{4}\right)\right|^2$
\\[8pt]
&
$\kappa=2$
&
$\frac{1}{8\sqrt{\pi}+5\pi}
\left[
-\tilde{K}_1(x)
+ 2\sqrt{2}\tilde{K}_{3/2}(x)
+2 \tilde{K}_2(x)
\right]$
&
$\frac{\sqrt{2}}{\left(8+5\sqrt{\pi}\right)\pi}
\left[
\left|\Gamma\left(\frac{3+ix}{4}\right)\right|^2
+2\left|\Gamma\left(\frac{5+ix}{4}\right)\right|^2
+\pi
\frac{x\sinh\left(\frac{\pi x}{4}\right)
+\cosh\left(\frac{\pi x}{4}\right)}{\cosh\left(\frac{\pi x}{2}\right)}
\right]$
\\[8pt]
&
\mr{2}{*}{$\kappa=10$}
& 
$\frac{1}{240\sqrt{\pi}+61\sqrt{5}\pi}
\left\{
20 \sqrt{2} \left[-\tilde{K}_{3/2}(x)+\tilde{K}_{5/2}(x)\right]
+\sqrt{5} \left[\tilde{K}_1(x)+4\tilde{K}_{3}(x)\right]
\right\}$
&
$
\frac{\sqrt{5}}{\sqrt{2} \pi\left(240+61\sqrt{5\pi}\right)}
\left[
\left(x^2+25\right)\left|\Gamma\left(\frac{3+ix}{4}\right)\right|^2
+40\left|\Gamma\left(\frac{5+ix}{4}\right)\right|^2
+\frac{10\pi}{\sqrt{5}}\frac{\left(x^2+5\right)\cosh\left(\frac{\pi x}{4}\right)
+4x\sinh\left(\frac{\pi x}{4}\right)}{\cosh\left(\frac{\pi x}{2}\right)}
\right]
$
\\[8pt]
\end{tabular}
\end{ruledtabular} 
\end{table*}
\end{turnpage}

Fig.~\ref{fig:DHO} reveals that the HF holes and correlation holes of $P(u)$ and $M(v)$ are surprisingly similar in size and shape.  It also shows that the introduction of correlation decreases the probabilities of $\omega \approx \pi/2$ and $x \approx 0$, indicating that the correlated electrons spend less time circularly orbiting their centre of mass.  This conclusion is supported by both the non-rigorous $\Upsilon(\omega)$ and $D(x)$ intracules and the rigorous $X(x)$ intracule.

However, the differences between $D(x)$ and $X(x)$ are significant.  $D(x)$ comes from the Wigner distribution and, as Eqs \eqref{Xkrhou} and \eqref{Dkrhou} show, it is the $O(\hbar)$ approximation to $X(x)$.  Its Fourier transform $\widehat{D}(k)$ decays as $k^{-3}$ for large $k$ and this creates a discontinuity in the second derivative $D''(x)$ at $x = 0$ \cite{Pars08, Boyd00}.  In contrast, $X(x)$ is smooth at $x = 0$.  One of the consequences of this misbehaviour at $x = 0$ is that, for the $\kappa = 2$ quantum dot, the Dot intracule's prediction $D(0) = 0.219$ overestimates the exact value $X(0) = 0.168$ by 30\%.

\section{\label{sec:He} Helium-like ions}
We now turn our attention to the helium-like ions. The Hamiltonian is obtained by substituting the harmonic potential $V(r)$ in \eqref{H0-QD} with the Coulombic potential 
\begin{equation}
	V(r)=-\frac{Z}{r},
\end{equation}
where $Z$ is the nuclear charge.  As before, the HF wave function and energy can be found by expanding the HF orbital in a Gaussian basis, optimizing both the coefficients and exponents.  We consider five values of $Z$, corresponding to the H$^{-}$, He, Li$^{+}$, B$^{3+}$ and Ne$^{8+}$ ions and, in this Section, we focus on their Position and Posmom intracules.

\subsection{Ground state}
\begin{table}
\caption{
\label{tab:EHeZ}
HF, radial and exact energies of various helium-like ions.}
\begin{ruledtabular} 
\begin{tabular}{llllll}
Atom							&\mc{1}{c}{H$^{-}$}		& \mc{1}{c}{He}			& \mc{1}{c}{Li$^{+}$}	& \mc{1}{c}{B$^{3+}$}	& \mc{1}{c}{Ne$^{8+}$}	\\
$Z$								&\mc{1}{c}{1}			&\mc{1}{c}{2}			&\mc{1}{c}{3}			&\mc{1}{c}{5}			&\mc{1}{c}{10}			\\
\hline	
HF								&$-0.487\,93$			&$-2.861\,67$			&$-7.236\,41$			&$-21.986\,2$			&$-93.861\,1$			\\
Radial							&$-0.514\,5$			&$-2.879\,0$			&$-7.252\,5$			&$-22.001\,5$			&$-93.875\,9$			\\
Exact							&$-0.527\,75$			&$-2.903\,72$			&$-7.279\,91$			&$-22.031\,0$			&$-93.906\,8$			\\
\hline
$E_{\text{c}}$ 					&$-0.039 82$			&$-0.042 05	$		&$-0.043 50$			&$-0.044 8$			&$-0.045 7$			\\
$\%E_{\text{c}}^{\text{rad}}$	&\mc{1}{c}{66.7}		&\mc{1}{c}{41.3}		&\mc{1}{c}{37.0}		&\mc{1}{c}{34.2}		&\mc{1}{c}{32.3}		\\
\end{tabular}
\end{ruledtabular} 
\end{table}

The HF orbital of the $^1S$ ground state was approximated by a Gaussian expansion \eqref{HF} with $N_\text{G} = 11$.  The exact wave function was approximated by the 64-term Hylleraas-type expansion \cite{Hylleraas29}
\begin{equation} \label{HeEx}
	\Psi(\bm{r}_1,\bm{r}_2)=\sum_{nlm}^3 c_{nlm}(r_1+r_2)^n(r_1-r_2)^{2l}r_{12}^m e^{-\alpha (r_1+r_2)}.
\end{equation}
We also considered the 64-term radially-correlated wave function \cite{ClementiEcR70, KogaEcR96}
\begin{equation} \label{HeRad}
	\Psi_{\text{rad}}(\bm{r}_1,\bm{r}_2) = \sum_{nl}^7 c_{nl}(r_1+r_2)^n(r_1-r_2)^{2l}e^{-\alpha (r_1+r_2)}.
\end{equation}

Table \ref{tab:EHeZ} gathers the HF, exact and radially correlated energies obtained from \eqref{HF}, \eqref{HeEx} and \eqref{HeRad}, respectively. Only the correct figures are reported \cite{HeZHF95, KogaEcR96, HeZEx07}, as well as the percentage of radial correlation ($\%E_{\text{c}}^{\text{rad}}$). As above, the correlation hole is defined as the difference between exact and HF intracules [Eq. \eqref{DI}]. We also define the radial and angular holes \cite{RadPu}
\begin{align}
	\Delta I_{\text{rad}}&=I_{\text{rad}}-I_{\text{HF}},\\
	\Delta I_{\text{ang}}&=I-I_{\text{rad}}.
\end{align}

\begin{figure}
	\subfigure[Hartree-Fock]{
	\includegraphics[width=0.4\textwidth]{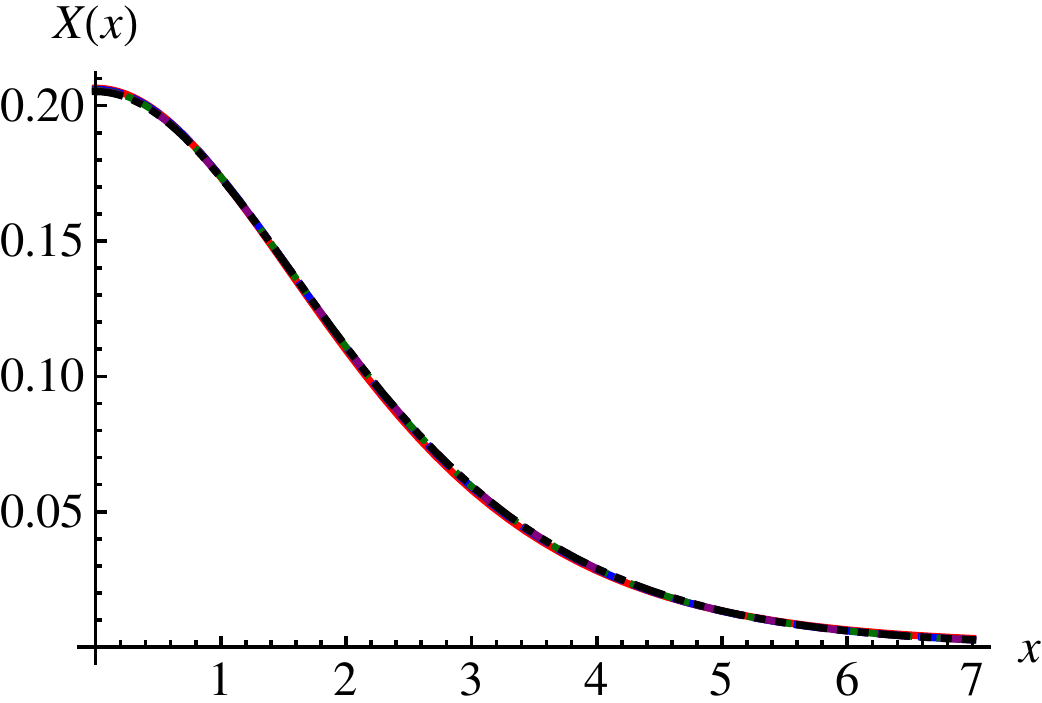}\label{fig:He_HFXx}}
	\subfigure[Exact]{
	\includegraphics[width=0.4\textwidth]{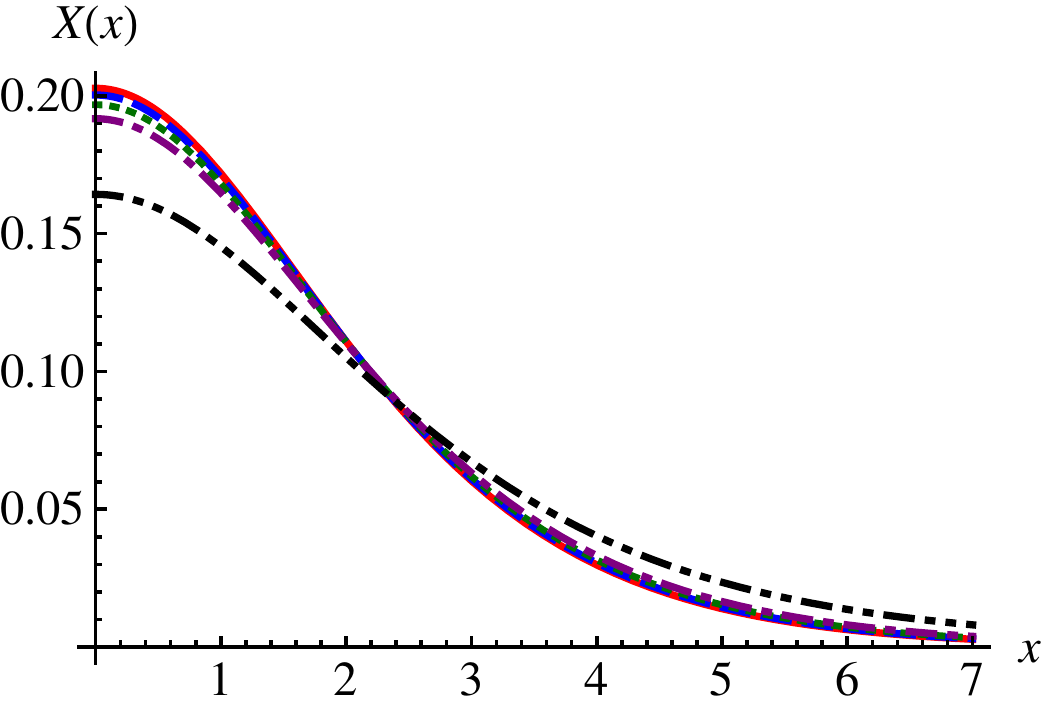}\label{fig:He_Xx}}
\caption{
Posmom intracules of helium-like ions: $Z=1$ (- $\cdot$ $\cdot$ -), $Z=2$ (- $\cdot$ -), $Z=3$ ($\cdots$), $Z=5$ (- - -) and $Z=10$ (---).}
\label{fig:He}
\end{figure}

Both $P(u)$ and $X(x)$ were obtained numerically and are shown in Fig.~\ref{fig:He}.  Increasing the nuclear charge barely affects $X_{\text{HF}}(x)$ because it produces an almost uniform contraction of the system.  However, the effect on the exact $X(x)$ is much larger and the values of $X(0)$ are 0.1642, 0.1917, 0.1968, 0.2003 and 0.2027 for $Z=$ 1, 2, 3, 5 and 10, respectively.  These values reveal that the electrons in H$^{-}$ spend less time mutually orbiting than those in Ne$^{8+}$.  This is consistent with the conventional view that H$^-$ is a more strongly correlated system than Ne$^{8+}$.

\begin{figure}
	\subfigure[Posmom]{
	\includegraphics[width=0.4\textwidth]{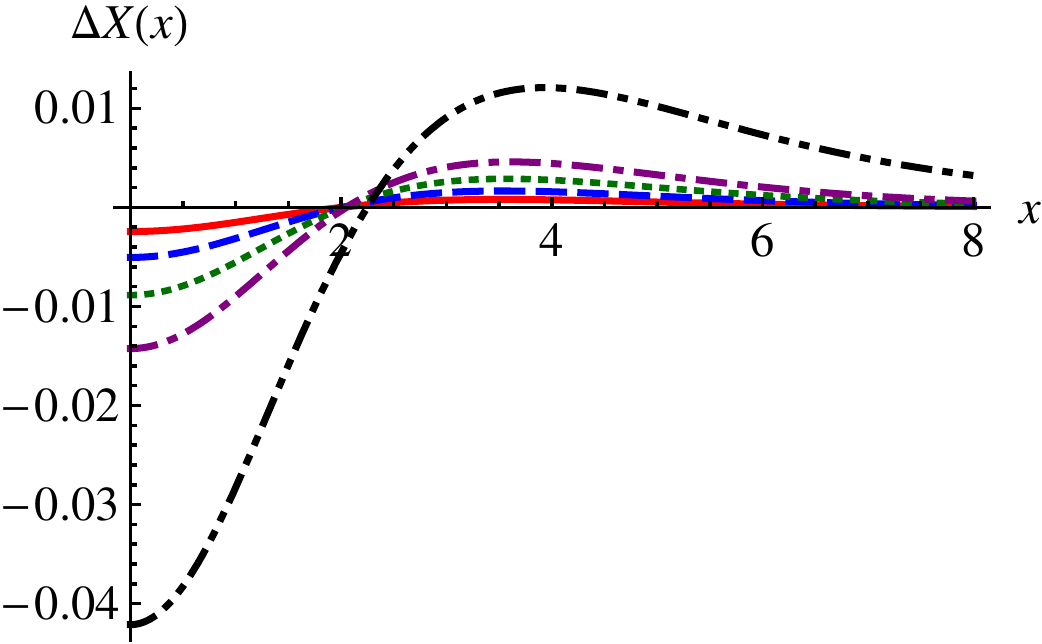}\label{fig:He_DXx}}
	\subfigure[Position]{
	\includegraphics[width=0.4\textwidth]{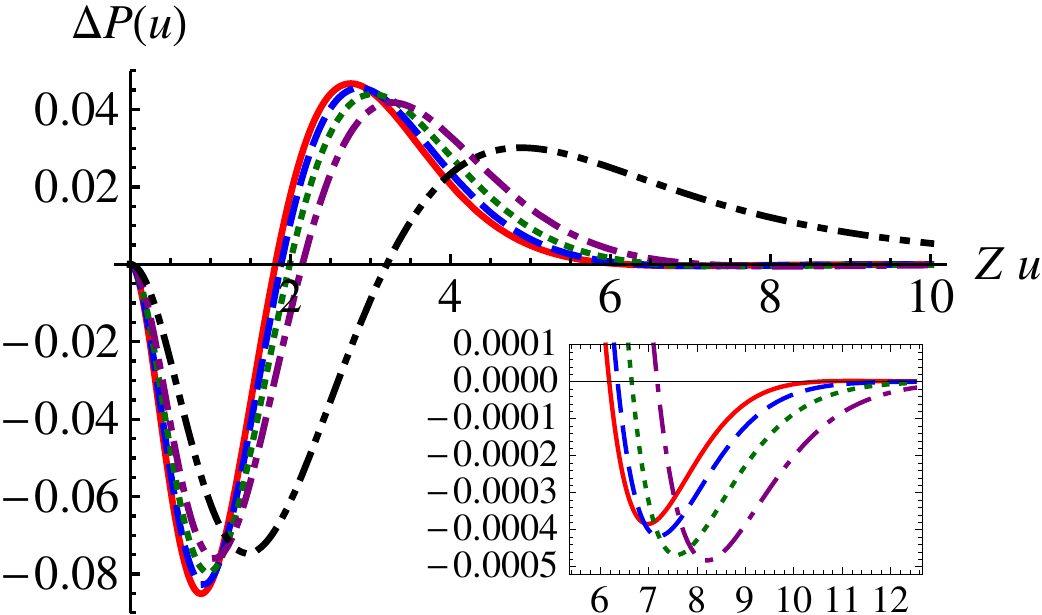}\label{fig:He_DPu}}
\caption{
Correlation holes of helium-like ions: 
$Z=1$ (- $\cdot$ $\cdot$ -), $Z=2$ (- $\cdot$ -), 
$Z=3$ ($\cdots$), $Z=5$ (- - -) and $Z=10$ (---).}
\label{fig:HeD}
\end{figure}

Fig.~\ref{fig:HeD} shows the Posmom and Position holes of the ions.  The depth of $\Delta X(x)$ decreases as $Z$ increases but the depth of $\Delta P(u)$ is almost constant as it is squeezed toward the origin. $\Delta P(u)$ exhibits a secondary hole, discussed in detail by Pearson \textit{et al.} \cite{Pearson09}, but this subtle correlation effect is not visitble in $\Delta X(x)$.

\begin{figure}
	\subfigure[Posmom radial correlation hole]{
	\includegraphics[width=0.38\textwidth]{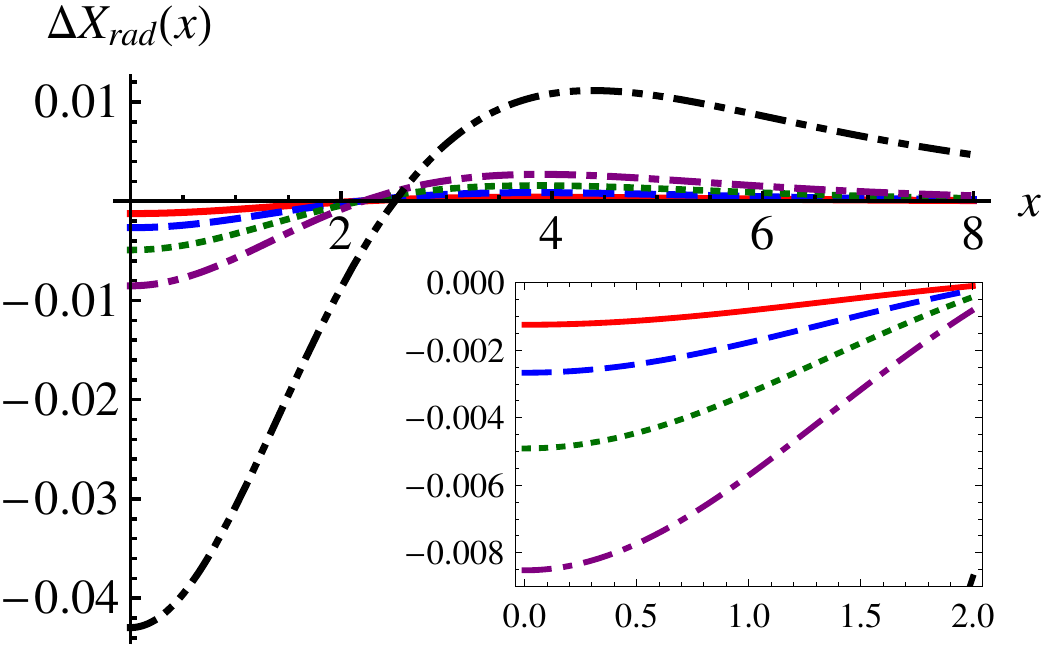}\label{fig:He_DradXx}}
	\subfigure[Posmom angular correlation hole]{
	\includegraphics[width=0.38\textwidth]{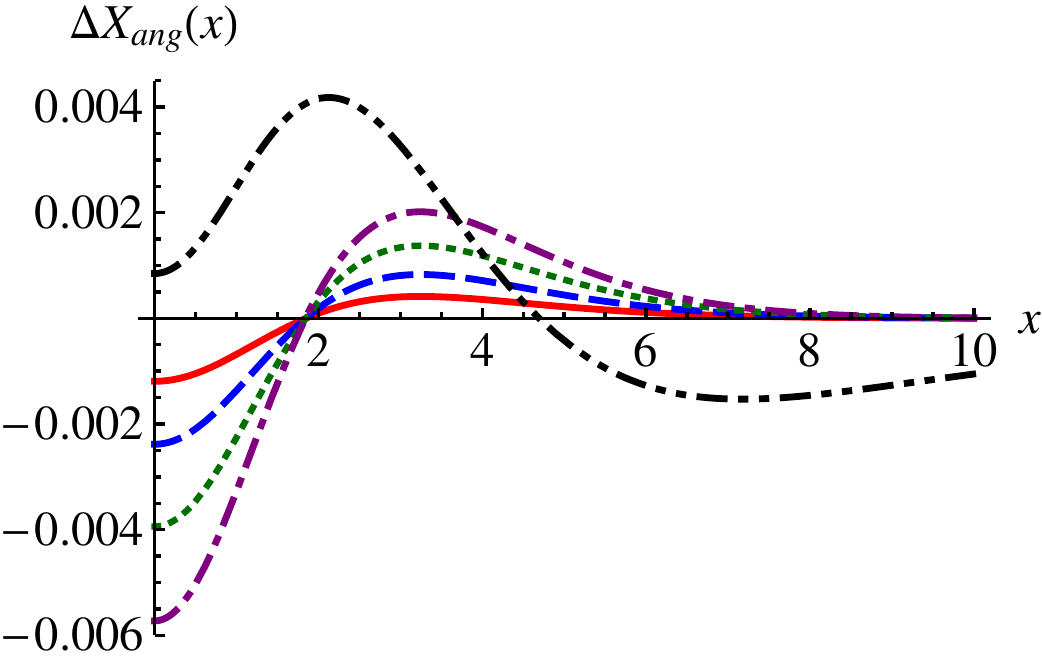}\label{fig:He_DangXx}}\\ 
	\subfigure[Position radial correlation hole]{
	\includegraphics[width=0.38\textwidth]{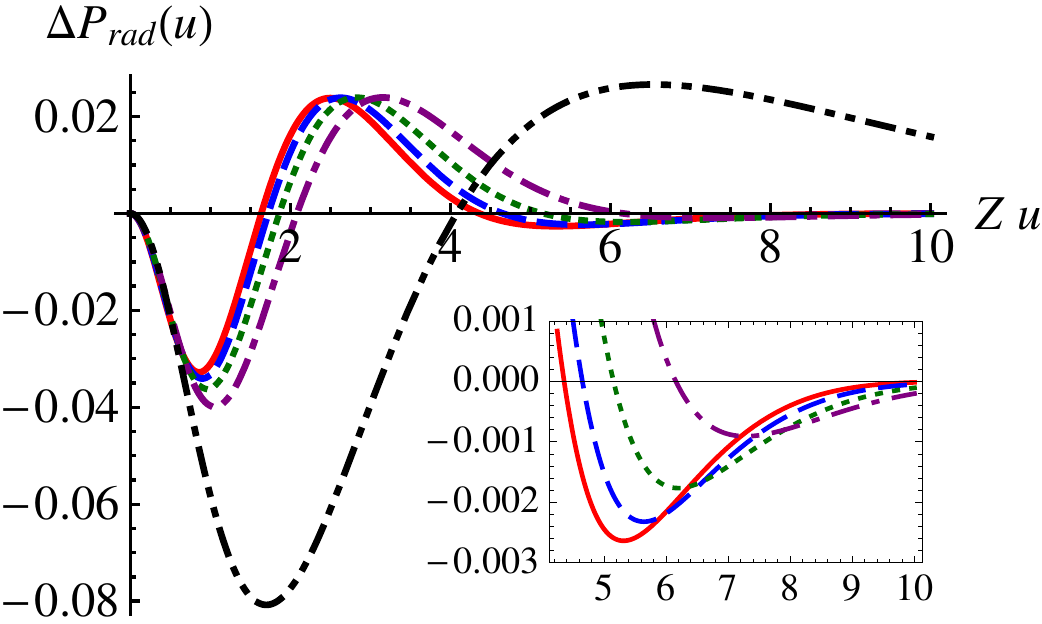}\label{fig:He_DradPu}}
	\subfigure[Position angular correlation hole]{
	\includegraphics[width=0.38\textwidth]{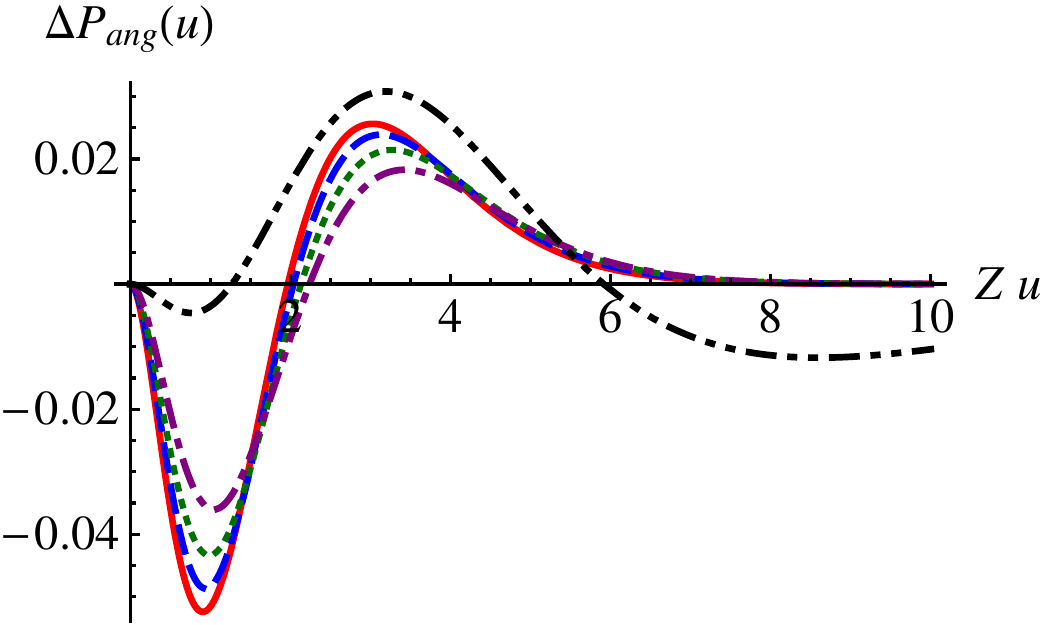}\label{fig:He_DangPu}}
\caption{
Radial and angular holes in He-like ions: $Z=1$ (- $\cdot$ $\cdot$ -), $Z=2$ (- $\cdot$ -), $Z=3$ ($\cdots$), $Z=5$ (- - -) and $Z=10$ (---).}
\label{fig:HeP}
\end{figure}

Radial correlation provides the majority (67\%) of the total correlation energy in H$^{-}$ but this decreases to 41\% in He, and to 32\% in Ne$^{8+}$, as angular correlation effects becomes dominant.  This shift is visible in $\Delta P_{\text{rad}}(u)$ and $\Delta P_{\text{ang}}(u)$, as shown in Fig.~\ref{fig:HeP}, but $\Delta X_{\text{rad}}(x)$ is always larger than $\Delta X_{\text{ang}}(x)$ and becomes almost identical for Ne$^{8+}$.

If we compare the insets in Fig.~\ref{fig:He_DPu} and Fig.~\ref{fig:He_DradPu}, we see that the \emph{radial} secondary hole is several times deeper than the \emph{total} secondary hole.  This implies that the radial secondary hole, which has been noted previously by Katriel \textit{et al.} \cite{RadPu}, is almost entirely cancelled by an angular secondary hole.

\subsection{Excited States}
\begin{table}
\caption{
\label{tab:X0He}
Origin Posmom intracule, $X(0)$, for various singlet and triplet excited states of the helium atom ($i,j=x,y,z$)}
\begin{ruledtabular} 
\begin{tabular}{ccc}
Configuration	& Singlet 		& Triplet 		\\
\hline
$1s^2$ 		& 0.2060		& --- 			\\
$1s2s$ 		& 0.0866		& 0.09520		\\
$1s2p$ 		& 0.0928 		& 0.09525		\\
$2s^2$ 		& 0.1647 		& --- 			\\
$2s2p$ 		& 0.1455 		& 0.1153 		\\
$2p_i^2$ 		& 0.1267 		& --- 			\\
$2p_i2p_j$ 	& 0.1567 		& 0.1538
\end{tabular}
\end{ruledtabular} 
\end{table}

\begin{figure}
	\includegraphics[width=0.4\textwidth]{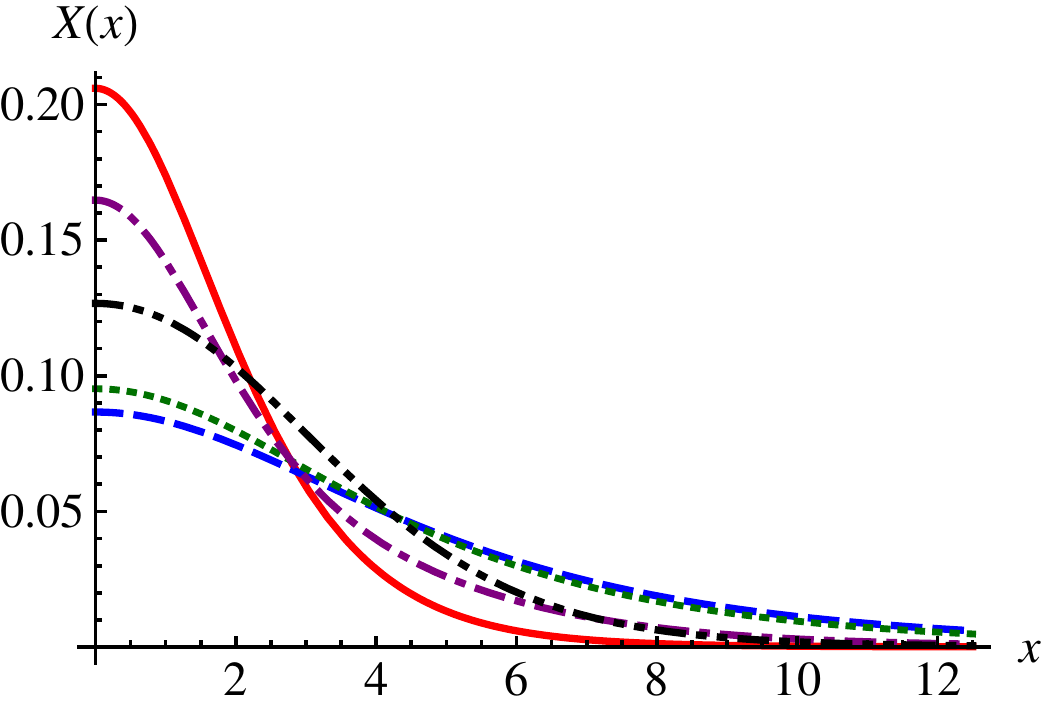}
\caption{
\label{fig:HeES}
Posmom intracule for states of the He atom: $1s^2\ {}^1S$ (---), $1s2s\ {}^1S$ (- - -), $1s2s\ {}^3S$ ($\cdots$), $2s^2\ {}^1S$ (- $\cdot$ -), $2p^2\ {}^1S$ (- $\cdot$ $\cdot$ -).}
\end{figure}

We have calculated the HF Posmom intracule for several excited states of the He atom using a Gaussian basis of 36 $s$-type functions with exponents $2^{-15},2^{-14},\cdots,2^{20}$ and 31 $p$-type functions with exponents $2^{-10},2^{-14},\cdots,2^{20}$. The maximum overlap method (MOM) has been employed for finding excited-state solutions to the HF self-consistent field equations \cite{MOM08}.

The intracules in the ground state ($1s^2\ {}^1S$) and four excited states ($1s2s\ {}^1S$, $1s2s\ {}^3S$, $2s^2\ {}^1S$ and $2p^2\ {}^1S$) are shown in Fig. \ref{fig:HeES}. Table \ref{tab:X0He} lists the values of $X(0)$ for these and other excited states.

When the electron pair occupies a more diffuse orbital, $X(x)$ becomes broader and $X(0)$ drops from 0.206 in the $1s^2$ state to $0.165$ in the $2s^2$ state, and to $0.127$ for the $2p^2$ state.  The decrease is even more marked when the electrons occupy orbitals in different shells, such as in the $1s2s\ {}^1S$ state where $X(0)=0.087$. However, if the two orbitals have the same principal quantum number, such as in the $2s2p \ {}^1P$ state, the decrease is smaller.

The Dot intracule has been calculated for the first excited state $1s2s\ {}^3S$ \cite{Pars08} and shows a small dip in $D(x)$ around $x = 0$, which we have previously attributed to the Fermi hole.  We now believe that that explanation was incorrect and that the dip is a failure of $D(x)$ to capture the behaviour of $X(x)$.

\subsection{D-dimensional helium atom}
\begin{figure}
	\includegraphics[width=0.4\textwidth]{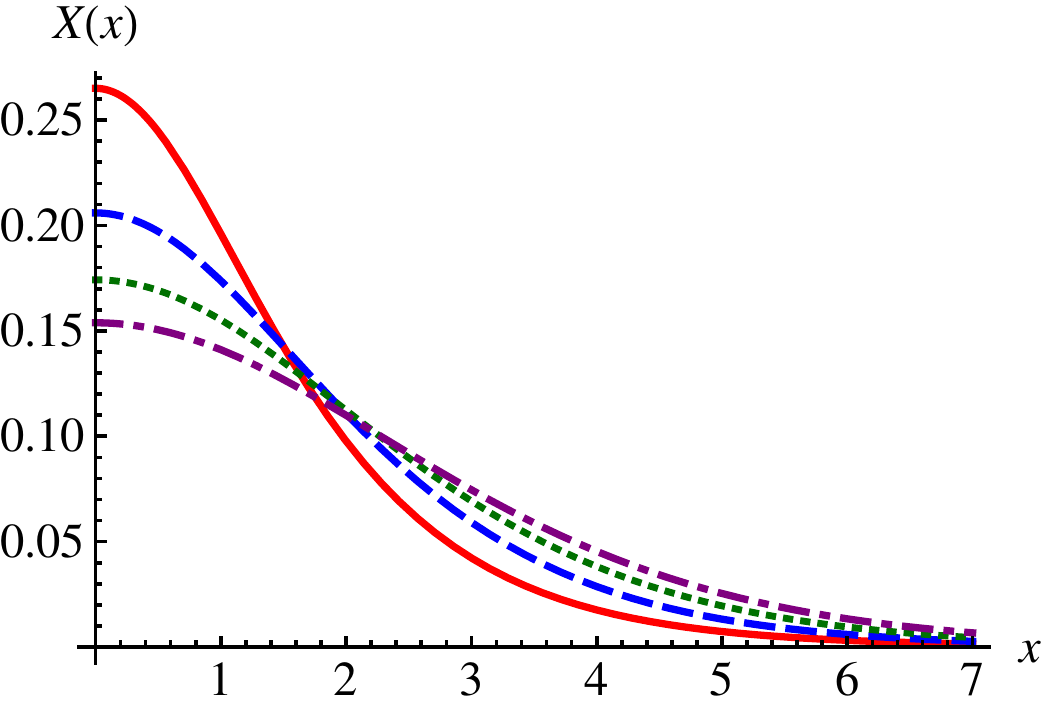}
\caption{
\label{fig:DHeXx}
HF Posmom intracules of the $\mathcal{D}$-dimensional He atom: $\mathcal{D}=2$ (---), $\mathcal{D}=3$ (- - -), $\mathcal{D}=4$ ($\cdots$), $\mathcal{D}=5$ (- $\cdot$ -).}
\end{figure}

Following the pioneering work of Loeser and Herschbach on the effect of dimensionality on the HF \cite{HerschbachDinter, HerschbachHFD} and exact energies \cite{HerschbachExD} of He-like ions, several other two-electron systems have recently been studied in $\mathcal{D}$ dimensions \cite{LoosHypSph09, LoosExcitSph, LoosHighDens09, LoosBall10, LoosInvarHigDen10, LoosTale2e10}.

The generalization of the Posmom intracule for a $\mathcal{D}$-dimensional space is straightforward. Eqs \eqref{XxFT}, \eqref{Xkrho2} and \eqref{Xkrhou} are unchanged and Eq.~\eqref{xkCssss} becomes
\begin{equation}
\label{xkCssssD}
	\left[ssss\right]_{\widehat{X}}=\frac{\pi^{\mathcal{D}}}{J^{\mathcal{D}/2}}.
\end{equation}
We used a large, even-tempered Gaussian basis, optimizing the coefficients $c_j$ to minimize the HF energy \cite{HerschbachHFD, LoosHighDens09}
\begin{equation}
\label{DEHF}
	E_\text{HF} = 
	2\int\psi_\text{HF}(r)\left[-\frac{\nabla^2}{2}+V(r)\right]\psi_\text{HF}(r)d\bm{r}
	+\iint\psi_\text{HF}^2(r_1)\left<\frac{1}{r_{12}}\right>_{\mathcal{D}}\psi_\text{HF}^2(r_2)d\bm{r}_1d\bm{r}_2,
\end{equation}
where 
\begin{gather}
	\left<\frac{1}{r_{12}}\right>_{\mathcal{D}} =
	\frac{F\left(\frac{3-\mathcal{D}}{2},\frac{1}{2},\frac{\mathcal{D}}{2},\frac{\text{min}(r_1,r_2)^2}{\text{max}(r_1,r_2)^2}\right)}{\text{max}(r_1,r_2)},
	\label{Coulomb-D}\\
	\nabla^2  = \frac{d^2}{dr^2}+\frac{\mathcal{D}-1}{r}\frac{d}{dr},
	\label{nabla}\\
	d\bm{r}=\frac{2\pi^{\mathcal{D}/2}}{\Gamma(\mathcal{D}/2)}r^{\mathcal{D}-1}dr,
	\label{Ddr}
\end{gather}
and $F$ is the Gauss hypergeometric function \cite{NISTbook}.  Our energies for $\mathcal{D}=$ 2, 3, 4 and 5 agree within a microhartree with the benchmark values of Herschbach and co-workers \cite{HerschbachHFD}.

Figure \ref{fig:DHeXx} shows how $X(x)$ changes with $\mathcal{D}$.  The observation that the intracule broadens as $\mathcal{D}$ increases is consistent with the conclusion of Herrick and Stillinger \cite{Herrick75} that the electrons in $\mathcal{D}$-helium can avoid each other more easily when $\mathcal{D}$ is large. Furthermore, they have shown that the binding energy of the ground state in $\mathcal{D}=5$ corresponds exactly to the binding energy of the $2p_i2p_j\ {}^3P$ state in $\mathcal{D}=3$. This feature is due to interdimensional degeneracies, first noticed by van Vleck \cite{vanVleck}, and observed for various systems \cite{HerrickJMP75, Herrick75, Doren86, Goodson91, DunnPRA1999, LoosExcitSph}.  We observe likewise that $X(x)$ for the $1s^2$ state in $\mathcal{D} = 5$ is identical to $X(x)$ for the $2p_i2p_j\ {}^3P$ state in $\mathcal{D} = 3$.

\section{Conclusion}
We have introduced a new two-particle density distribution, the Posmom intracule, which condenses information about both the relative position and relative momentum of the particles.  We have shown how to construct this distribution from the many-particle wave function and we have shown that the Dot intracule $D(x)$ in a first-order approximation of the Posmom intracule $X(x)$.  We have applied our new formalism to two-electron quantum dots and the the He-like ions.  A comparison between various intracules (Position, Momentum, Angle, Dot and Posmom) has been carried out, showing the interrelated information conveyed by these two-particle probability distributions.

\begin{acknowledgments}
We thank Joshua Hollett for helpful comments.  YAB thanks the ANU Research School of Chemistry for a PhD scholarship. PMWG thanks the Australian Research Council (Grants DP0984806, DP1094170, and DP120104740) for funding. PFL thanks the Australian Research Council for a Discovery Early Career Researcher Award (Grant DE130101441). PFL and PMWG thank the NCI National Facility for a generous grant of supercomputer time.
\end{acknowledgments}

\end{document}